# Software-Defined Radio GNSS Instrumentation for Spoofing Mitigation: A Review and a Case Study

Erick Schmidt, *Student Member, IEEE*, Zach A. Ruble, *Member, IEEE*, David Akopian, *Senior Member, IEEE,* Daniel J. Pack, *Senior Member, IEEE*

*Abstract*— Recently, several global navigation satellite systems (GNSS) emerged following the transformative technology impact of the first GNSS - US Global Positioning System (GPS). The power level of GNSS signals as measured at the earth's surface is below the noise floor and is consequently vulnerable against interference. Spoofers are smart GNSS-like interferers, which mislead the receivers into generating false position and time information. While many spoofing mitigation techniques exist, spoofers are continually evolving, producing a cycle of new spoofing attacks and counter-measures against them. Thus, upgradability of receivers becomes an important advantage for maintaining their immunity against spoofing. Software-defined radio (SDR) implementations of a GPS receiver address such flexibility but are challenged by demanding computational requirements of both GNSS signal processing and spoofing mitigation. Therefore, this paper reviews reported SDRs in the context of instrumentation capabilities for both conventional and spoofing mitigation modes. This separation is necessitated by significantly increased computational loads when in spoofing domain. This is demonstrated by a case study budget analysis.

## I. INTRODUCTION

THE success of the US Global Position System (GPS) promoted broader deployment of other global navigation satellite systems (GNSS) as well for providing position, velocity and time (PVT) information for users in civil, commercial and military applications [1], [2]. After GPS became available to commercial and civil markets, many components of critical infrastructure and broadly used applications started relying on the continuous availability of PVT information. Today, a sudden shutdown of GNSS would have a tremendous impact on systems relying on them for navigation or timing. Examples of GNSS-reliant ecosystems include conventional and emerging autonomous transportation [3], cellular networks [4], and even regulation and measurement of phasors in power systems [5], [6].

With so many existing and new markets depend on GPS, it is essential for GPS receivers to withstand interference, intentional or otherwise. GPS signals can experience unintentional interference from other radio frequency (RF) emitters or intentional interference due to jamming or spoofing attacks. Spoofers are intelligent jammers that transmit specific counterfeit GNSS-like signals to force the receiver to compute erroneous positioning and timing [7], [8].

There have been many techniques developed to deal with both intentional and unintentional interference over the years. In particular, arrays of antennas are effective in validating known direction-of-arrivals (DOA) of satellite signals at the expense of increasing receiver cost and size [9]-[11], which is undesirable for mass-market applications. There exist many single antenna techniques as well, such as [7], [8], [12]-[16].

Nevertheless, spoofing techniques continuously evolve, and there are no universal mitigation techniques that address all current and future threats. In that light, upgradability of the GNSS receivers becomes an important issue to support current functionality for best achievable overall performance and protection against spoofing attacks. Conventional GNSS receivers rely heavily on hardware (HW) components due to intense computational requirements and are not flexible for essential upgrades. Emerging software-defined radio (SDR) solutions implement most of the critical operations in software (SW) mode and are highly flexible for upgradability. The flexibility introduced by SDRs also makes it preferred research and development instrumentation for fast prototyping and testing of new receiver architectures and algorithms.

Still, state-of-the-art SDR solutions do not match computational power of HW-based receivers [17], [18], and often do not support real-time operations. For acceleration, SDRs often employ field programmable gate arrays (FPGAs) [19]-[27], digital signal processors (DSPs) [28]-[31] or are fully implemented in SW on a host PC [32]-[54]. Of the mentioned SDR GNSS receivers, only a few address interference mitigation techniques.

Protection against spoofing attacks requires significant additional computational resources, which are not guaranteed by reported general-purpose solutions. With an existing

Manuscript first submitted on March 27, 2018 to the IEEE Transactions on Instrumentation and Measurement.
E. Schmidt, and D. Akopian are with the Department of Electrical and Computer Engineering, The University of Texas at San Antonio, San Antonio, TX 78249 USA (e-mail: erickschmidtt@gmail.com; david.akopian@utsa.edu).

Z. A. Ruble, and D. J. Pack are with the Department of Electrical Engineering, University of Tennessee at Chattanooga, Chattanooga, TN 37403 USA (e-mail: zachary-ruble@utc.edu; daniel-pack@utc.edu).

considerable research spectrum related to the general field of GNSS SDRs [43], [45], [55], [56], the specific domain of spoofing mitigation in the context of instrumentation capabilities is not systematically reviewed to the best of the authors' knowledge, and this paper addresses this existing gap. Therefore, this paper, first, contributes a general-to-specific SDR review through the lens of interference mitigation and instrumentation budgeting. It includes: (1) an overall GNSS receiver classification, (2) reported performance of SDRs for this application, and (3) a demonstration of a significant surge in complexity between conventional GNSS and advanced mitigation operational modes by using a case study receiver and a cross-correlation mitigation algorithm.

Section II presents a state-of-the-art overview of general domain GNSS-SDR solutions. Section III examines reported SDR applications for spoofing mitigation domain in terms of instrumentation capabilities. Section IV presents a case study receiver computational budget of common operations. Section V presents a reduced complexity minimum mean squared error (MMSE) technique proposed in [12], which is used as a case study integrated with a receiver previously developed by the authors. Section VI examines aforementioned MMSE implementation in a fast prototyping real-time SDR testbed, along with a computational complexity budget analysis. Section VII presents simulation and performance results of mitigation algorithms to validate functionality. Section IX finalizes with concluding remarks.

## II. GNSS Receiver Design and Overview

Software-defined radio emerged with a substantially beneficial purpose: to place the analog-to-digital converter (ADC) as close to the antenna front-end (FE) as possible, so that all samples from the ADC are post-processed in a reconfigurable SW mode rather than HW. This adds configurability, flexibility, and upgradability [55]. This trend has gained attention in GNSS receivers; especially and recently, in real-time operation capabilities.

It was not until the last two decades that general-purpose processors (GPPs) have gained enough processing power to achieve real-time operations that previous HW application-specific integrated circuits (ASICs) were performing, such as HW correlators. The main components of a GNSS receiver are an RF block, which consists of an active antenna, a low-noise amplifier, and a FE, followed by three common baseband blocks: acquisition, tracking, and navigation.

Acquisition is a process of coarse synchronization of the received signals with locally-generated correlating replicas for estimating their time and frequency misalignment. For acquisition, recent advances use FFT-based techniques to either replace correlation operations in the frequency domain (carrier replica) or the time (code replica) domain. Other advanced acquisition algorithms use shared FFT techniques for joint-space searching in both domains for even faster computations [57].

Tracking is a process of fine synchronization where continuous alignment of the incoming signal with so-called local pseudorandom (PRN) code replicas (time domain) and carrier replicas (frequency domain) need to be attained. This becomes the most challenging task in GPS SDRs since tracking loops conduct continuous processing of an incoming satellite signal by estimating carrier phase and code phase offsets for a stable synchronization lock. This is done for all previously acquired (visible) satellite signals. Code phase estimation is obtained by a delay-locked loop and carrier phase estimation is obtained by a phase-locked loop and/or a frequency-locked loop [32], [47], [58]. This allows for fine synchronization in both the time and frequency domains, resulting in successful despreading of the incoming GNSS signals for navigation data extraction. During tracking synchronization, correlator outputs from the delay-locked loop are grouped into unique frames to form navigation data and to compute pseudo-range measurements [32], [47], [55].

Finally, if navigation data is successfully collected from enough satellites (four is the minimum), then the tracking stage gathers data from all channels, aligns data in a systematic set, and runs navigation algorithms to solve user PVT solutions [47], [58].

GNSS-SDR receivers are designed based on the aforementioned baseband modules and the RF block. Authors in [55], [56] describe SW radio receivers with various HW/SW configurable components where baseband functionality (i.e., acquisition, tracking, and navigation) can be distributed after the RF block (FE sampler): reconfigurable hardware such as FPGAs, coprocessor units such as graphical processing units (GPUs), embedded specialized GPPs such as DSPs, and general-purpose host PCs. Authors in [55] have defined three architectures that divide GNSS functionality onto these components: *Classical HW, Hybrid*, and *Fully SW* architectures, based on where most of these baseband modules are distributed. Other authors distinguish SDR categories between post-processing and real-time solutions [56], based on SDR being real-time or not, and whether it is implemented on a PC or an embedded system.

As an alternate approach, we focus on correlators as being the most computational consuming operation in the receiver. Additionally, while some authors associate FPGAs and DSPs as being in the same category [45], [55], other authors distinguish DSPs for their *ease-of-use* [59] and for having a non-HW-configurable GPP. Therefore, we categorize FPGAs and DSPs separately. We also isolate DSP-based GPPs from PC GPPs. In addition, by considering aforementioned HW/SW configurable components by their *ease-of-use*, we define a category similar to *Fully SW*, which includes an even higher-level component recently adopted in SDR receivers: *a prototyping software (P-SW) platform*. A *P-SW platform* is defined as a high-level programming framework working atop the host PC operating system (OS) and has the ability to efficiently manage and optimize PC computational capabilities such as parallelism and multi-threading (MT). Such platforms or environments can be *LabVIEW* (LV), *MATLAB + Simulink* (M+S), or even open-source solutions such as *GNU Radio* [60] and *Python*. Many of these *P-SW platforms* offer *ease-of-use* and fast prototyping by means of built-in block implementations. We then define the following categories: (1) *FPGA* where all correlations occur, and a host PC with (optional) PVT solutions and pre-configurations. (2) *DSPs* or embedded GPPs, and a host PC for (optional) PVT solutions and/or output visualizations. (3) *Host PC* where most baseband functions are executed. Finally, (4) *P-SW platform*, for fast

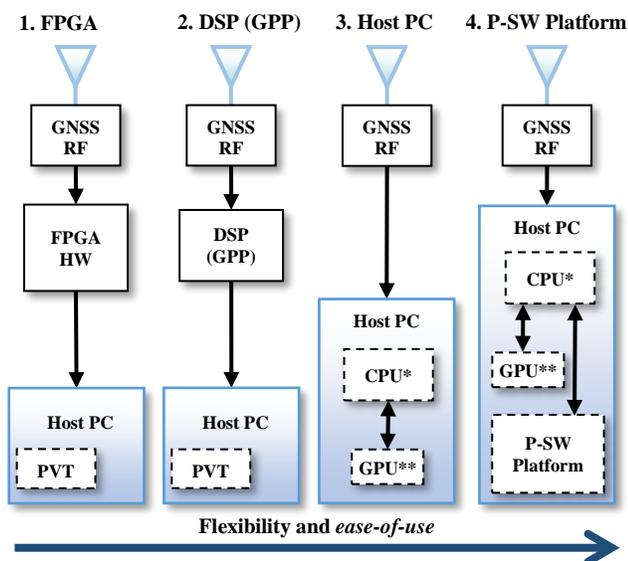

Fig. 1. GNSS SDR categorization based on flexibility and *ease-of-use*.

*C/C++, SIMD, OpenMP, MT, etc.
**GPU API libraries such as CuFFT.

TABLE I
REPORTED GPS RECEIVERS IN SDR CATEGORIES (1)-(4) (SEE FIG. 1)

| SDR category | Real-time capability | SW/HW architecture | Acceleration factors |
|---|---|---|---|
| *1. FPGA* [19]-[27] | Non-real-time [24], real-time [19]-[23], both [25]-[27] | FPGA/DSP board [19]-[21], [25], GPP-based PVT [19]-[23], [25]-[27], host PC visuals [22]-[24], [26] | FPGA accelerators [19]-[27] |
| *2. DSP (GPP)* [28]-[31] | Real-time [28]-[31] | DSP board [28]-[31], host PC visuals [28], [30], [31] | DSP accelerators [28]-[31], bit-wise operations [29], [31] |
| *3. Host PC* [32]-[45] | Non-real-time [32]-[34], real-time [35]-[43], both [44], [45] | C++-based [32]-[35], [38]-[41], [43], [44] APIs [34], [38], [41], [43], [44] Linux OS [36], [37], [45] GPU libraries [34], [39], [40], [41], [43] GUI [38], [41], [43], [44], [45] | Optimized libraries [32], [35], [38], [39], [41], [43] SIMD [33], [34], [39]-[41], [43], [45] MT [33]-[35], [38], [41], [43], [45] assembler [36], [38], [39], [44], [45] GPU APIs [33], [34], [39]-[41], [43] RT-Linux [36], [37], [45] bit-wise operations [35], [36], [41]-[43] OpenMP API [38], [41], [43] |
| *4. P-SW Platform* [46]-[54] | Non-real-time [46]-[51], both [52]-[54] | MATLAB-based [47]-[49], M+S [50], LV-based [46], [53], [54], open-source [51], [52] | FPGA accelerators [46], optimized libraries [52]-[54], MT [52]-[54], SIMD [52]-[54], P-SW accelerators [46], [53], [54] |

prototyping. This final category is a novelty in this paper and offering the highest flexibility in terms of SW configurability and *ease-of-use*. We attempt to categorize currently reported SDR solutions in said categories as seen in Fig. 1, with optional hybrid approaches across. We also avoid ASIC solutions [17], [18] as we focus on purely SDR implementations.

Table I summarizes current reported SDR solutions organized according to listed categories (1)-(4) (see Fig. 1). For a given receiver category (row), all receiver references falling into this category are listed under said section. Additionally, receiver category references are divided into three other columns: real-time capabilities, their overall SW/HW architecture, and leveraged factors for acceleration, if any. These three rightmost columns are independent of each other, and the acceleration factor column can repeat a receiver reference based on implemented acceleration techniques.

Of all these SDR solutions, the combination of real-time and *ease-of-use* implementations are considered the most valuable for research. Many solutions in category (1) implement reconfigurable HW accelerators in correlators [19]-[27] but offer less flexibility and *ease-of-use*, as opposed to GPPs found in categories (2)-(4) [59]. Category (2) offers low-power embedded solutions such as DSP boards [28]-[31], which are not tailored for optimal speed. The *ease-of-use* of this category is considered intermediate to advanced since it often requires low-level programming such as bit-wise operations [29], [31]. Category (3) offers more flexible environments, as it can use more generic programming skills and optimized libraries in the form of application programming interfaces (APIs) such as fast Fourier transforms (FFTs), [32], [35], [38], [39], [41], [43]. Many accelerators in this category are C/C++-based and exploit multicore abilities for parallelism [59], MT [33]-[35], [38], [41], [43], [45], single-input multiple-data (SIMD) instructions [33], [34], [39]-[41], [43], [45], OpenMP APIs to schedule CPU multicore resources with noticeable speedups on correlators [38], [41], [43], [59], GPU-based APIs for parallel arithmetic operations [33], [34], [39]-[41], [43], real-time (RT) Linux OS [36], [37], [45], among others (see Table I). Category (3) can be associated with the *Fully SW* option mentioned in [55]. Nevertheless, category (3) receivers often require advanced dedicated software solutions which have evolved during many years and optimizations using dedicated OS platforms such as RT Linux [43], [44], assembler instructions [36], [38], [39], [44], [45], and so on. Moreover, API-based SDRs typically do not give access to the source code [41], thus, limiting tight integrations with potential research algorithms. Category (4), which uses a *P-SW platform* for fast prototyping, commonly offers an *ease-of-use* framework such as in LV-based [46], [53], [54], and M+S-based [50]. This property enables researchers to test algorithms quickly and generally requires less programming effort, and time, e.g., authors [51] developed a Python-based vector tracking multi-receiver. Nonetheless, SW platform-based SDRs often lack accelerators due to their *ease-of-use* nature. Therefore, a scarcity of real-time receivers in said category proposes a challenge for ongoing research.

Recent research has proposed accelerators for category (4) that offer a combination of real-time and fast prototyping [53], [54]. In terms of interference mitigation in the context of instrumentation capabilities, said SDR fits well for research extensions.

## III. REVIEW OF GNSS SDRs FOR INTERFERENCE MITIGATION

Introducing algorithms that address interference mitigation generally increase the computational complexity of the receiver; thus, only a few of the many existing GNSS-SDR approaches operate in real-time. For this reason, this section separates this category of receivers. Furthermore, we examine reported interference mitigation techniques with their corresponding host receivers.

### A. GNSS SDRs with interference mitigation

Interference and spoofing of GNSS signals is a vast area of research. As mentioned in [7], certain signal properties of GNSS signals are vulnerable to and exploited by interference and spoofing attacks. An overview of types of interference attacks can be categorized into *jamming* and spoofing, which can be very similar but should be distinguished from one another. Some forms of jamming include narrowband or wideband continuous wave and RF interference (RFI) often called *chirping* signals [10], [14]. These techniques aim to make the receiver lose lock by transmitting overpowering signals. For spoofing, the attacks rely on subtlety by trying to take over the receiver's current position or time, rather than blocking its signal altogether. One of the most common spoofing approaches is meaconing, which is based on relaying satellite signals with increased power to introduce a delay on the target receiver and influence PVT outputs. The reader is directed to [7], [8], for more relevant information on jamming and spoofing attacks.

Table II presents a categorized set of SDR solutions found in the literature that use interference mitigation techniques based on mitigation types described in [8], as well as their reported SDR platform and real-time capability. Based on the *mitigation categories* column, authors in [8] described these four groups of interference mitigation: 1) *signal processing-based*, which exploit stages of a GNSS receiver on the RF chain, i.e., automatic gain control monitoring, as well as other common GPS stages such as acquisition, tracking, and/or navigation, 2) *cryptographic-based*, which attempt to provide encryption on GNSS signals, specifically on payload navigation data broadcast from the satellites for additional layers of security and authentication, 3) *correlation with other GNSS signals*, which utilize additional constellations, frequency bands, and/or sensors to monitor and detect counterfeit signals on current GNSS signals, and 4) *radio spectrum and antenna-based*, which exploit multiple antenna techniques, such as angle-of-arrival (AOA), to spatially pinpoint the counterfeit signal sources for dissolution from authentic signals.

Additionally, the *specific technique* column lists keywords that describe the overall method utilized for each mitigation category. For each overall method referenced in the specific technique column, an associated *SDR platform* is also referenced as the solution used, listed in the same order used in the specific technique column. If no SDR was reported for the specific technique, an *N/A* is used. For the *real-time* column, the reference correlates to the specific technique's reference, and not the SDR platform's reference.

#### 1) Signal processing-based

For the *signal processing* mitigation category found in Table II, several authors modify a certain step of the GNSS receiver chain to assess and mitigate interference. Similarly, they employ digital signal techniques for post-processing and analysis. In [61], authors modified a real-time SW receiver for detecting the presence of spoofing based on live post-correlator outputs from the tracking stage of their GNSS receiver, *NGene* [45]. These correlator outputs were filtered through a Chi-square statistic to detect anomalies. The *NGene* receiver successfully attained live spoofing detection by using two additional correlators per channel. Similarly, authors in [62] modeled correlator outputs with and without spoofing as a Gaussian distribution and attempted to detect spoofing by monitoring the model variance. Their simulations were run on a modified version of the *GSNRx* receiver [34]. Another post-correlation application was introduced in [63], where authors tracked vicinity peaks of the acquisition output for possible spoofing signals. This technique was implemented on a real-time open-source receiver, *GNSS-SDR* [52]. Authors in [64] evaluated the performance of wavelets for radio frequency interference (RFI), and notch filtering for continuous wave interference using recorded GPS data on their SW receiver platform *IpexSR* [38]. Vector tracking correlators reported in [65] rely on joint vector processing of concurrent tracking channels for monitoring of possible attacks and discards pseudo-range measurements from suspected nefarious channels before calculating the PVT solution. This solution used the aforementioned *GSNRx* platform in offline mode. A non-real-time Python-based vector tracking receiver, *PyGNSS* [51], applied direct GPS positioning (DP) [66] and direct timing estimation (DTE) [67] techniques, which are based on a maximum likelihood estimator (MLE), to discern spoofed GPS parameters in vector processing, such as clock bias. Power analysis and monitoring were reported in [62] with *GSNRx* in offline mode. Finally, other signal processing techniques used cross-correlation properties of PRN codes in optimization techniques, such as MMSE [12], [68], and multi-stage Weiner filtering, to separate the spoofer from authentic signals [69].

TABLE II
OVERVIEW OF SDR SOLUTIONS WITH INTERFERENCE MITIGATION TECHNIQUES DIVIDED IN CATEGORIES [8]

| Mitigation category | Specific technique | SDR platform | Real-time? |
|---|---|---|---|
| Signal processing | Post-correlators: [61], [62], [63] | [45], [34], [52] | [61], [63] |
| | wavelet + notch filter [64] | [38] | N/A |
| | vector-tracking correlators [65] | [34] | N/A |
| | MLE based on DP [66], and DTE [67] | [51], [51] | N/A |
| | Power analysis [62] | [34] | N/A |
| | MMSE + MWF [68], [69] | [53], N/A | [68] |
| Cryptograph | SCER [70], [72] | [71] | [72] |
| Correlation w/ other GNSS signals | GPS C/A and P(Y) correlation [73], [74], [75], [76], [77] | [71], [74], [29], [29], [51] | [73], [76] |
| Radio spectrum and antenna | Space-time correlation [79] | [34] | N/A |
| | AOA [80], [81] | [80], [71] | [80] |
| | Antenna arrays [82], [83], [84], [85], [86], [87] | N/A, [34], [34], [52], [52], [52] | [87] |

*2) Cryptographic-based*

For *cryptographic-based* techniques, authors in [70] used an improved *GRID* receiver [29] with an anti-spoofing and defender-receiver testbed reported in [71] for cryptographic mitigation techniques based on security code estimation and replay (SCER). The attack obtained an estimate of the security chip used on the received encrypted signals to replay it for authentication. The testbed was able to detect and mitigate SCER attacks in offline mode by using the modified GRID testbed [71]. In [72], authors achieved real-time mitigation for up to 14 channels with said attacks.

*3) Correlation with other GNSS signals*

In *correlation with other GNSS signals*, authors in [73]-[77], proposed using the GPS L1 band, which contains both the civilian C/A and the encrypted military P(Y) signals in the in-phase and quadrature, respectively. Authors correlated code phase and timing relations between both codes to detect possible spoofers. Receivers used were variations of *GRID* for [73]-[76], and *PyGNSS* for [77]. For most offline receivers, an existing Texas Spoofing Test Battery (TEXBAT) recording files database [78], which provide common spoofing scenarios for static and dynamic attacks, was used.

*4) Radio spectrum and antenna-based*

Finally, for the *radio spectrum and antenna-based* category, several techniques, such as space-time correlation introduced in [79], were proposed for multi-antenna testbeds. Spoofing and jamming signals were also detected by DOA and AOA of incoming satellite signals in [80]-[86]. However, the cost and additional equipment needed for such a receiver may not be suitable for all applications. Authors from [87] implemented an antenna array anti-spoofing testbed in their real-time *GNSS-SDR* receiver [52].

While many SDR receivers with interference mitigation integration were listed in Table II, only [61], [63], [68], [70], [71], [73], [76], [80], [87] provided real-time mitigation capability. Said capability is considered state-of-the-art because of its increased computational complexity in spoofing domain.

*B. Interference mitigation solutions comparison*

As an attempt to narrow the aspects of instrumentation capabilities on previously-discussed SDRs with interference mitigation, in this subsection, we include four selected popular receivers for a more detailed analysis. Table III provides this summary. It also includes a receiver from previous work [53], [54], for a demonstration of computational complexity distribution of a case study interference mitigation algorithm (Section VI). Three of five solutions are category (4) *P-SW platform* with two being real-time: LV-based solution [53], [54], and *GNU Radio*-based solution [52]. Another real-time solution is category (2) *DSP* from Humphreys et al. *GRID* [29], [71]. Several maximum sampling rates are seen across the reported and cited references, ranging from 5 Msps to 25 Msps (see Table III). As for real-time tracking, between 1 and 22 channels have been reported for selected solutions. When using 22 channels, typically half are used for conventional satellite tracking and the other half for interference injection and mitigation [71].

*1) Jafarnia-Jahromi et al. (GNSRx)*

Jafarnia-Jahromi et al. have several research papers concerning their *GNSRx* [34] receiver. They implemented signal processing-based interference mitigation techniques, as well as radio spectrum and antenna-based methods (see Table III). Since their SDR receiver is not real-time, a Spirent GPS simulator was used to generate signals for testing. The receiver uses a C++ modular design and uses 3 channels from a NI FE. They successfully implemented the receiver with 22 channels and sampling rates up to 25 Msps, normal and spoofed.

In the receiver, various GNSS signal operations are divided into high, medium, and low rate computational categories, which are performed in the 4-50 MHz, 50-1000 Hz, and 20 Hz or less range, respectively. To make the system adaptable to new algorithms, a modular object-oriented approach is used. The Doppler removal and correlation (DRC) object, used to track a given satellite signal, incurs the largest computational burden. To improve DRC processing, SIMD instructions for x86 processors and a multi-threaded architecture are implemented. These are used as C++-based accelerators for faster computations.

Additionally, an NVIDIA 8800GTX GPU is used in *GSNRx* to leverage DRC operations. The high degree of parallelism, as a result of the large number of available GPU threads, provides considerable processing improvements. GPU co-processors can be divided into small threads. Each thread computes the local code and carrier phase, Doppler removal, and local code multiplication for each sample it is responsible for and sums the result. It was shown in [34], that the average DRC processing time for 1ms of data on eight satellites was less than 1 ms when sampled at 25 Msps when using the GPU, providing a real-time operation, which was implemented in [40].

In [79], [83], [84], antenna techniques were used with *GSNRx* to achieve spoofing detection and immunity. In [79], space-time correlation was used and in [83], [84], an array of antennas was implemented to detect and mitigate spoofing attacks. In [62] and [65] they implemented vector tracking and power analysis mitigation techniques, respectively. None of these testbeds were reported as real-time since they used recordings from the Spirent simulator and used real, recorded signals to assess their performance.

*2) Humphreys et al. (GRID)*

Authors in [71] developed an augmented version of *GRID* [29] for real-time spoofing and detection of signals. The conventional *GRID* [29] receiver can track hundreds of GNSS live channels in its latest reported iteration [59], nonetheless, when spoofing is implemented, this computational power decreases. The augmented system [71] is capable of tracking up to 12 live authentic channels and 12 spoofed channels (see Table III). The proposed SDR uses a DSP, with most code written in C++ for upgradability. Most of the optimizations carried out in GRID are bit-wise parallel operations leveraged from the DSP's architecture by using built-in AND, NOR and XOR modules. They also use look-up tables stored in memory for fast local carrier and code generation. Their receiver's FE was able to sample at 5.714 MHz with an IF of 1.405 MHz. Its full range capability is a combination receiver and spoofer testbed, which can simultaneously generate 12 spoofer channels and defend them as reported in their latest SCER attack iteration [72].

TABLE III
COMPARISON TABLE FOR DIFFERENT SDR PLATFORMS ON INSTRUMENTATION CAPABILITIES FOR INTERFERENCE MITIGATION TECHNIQUES AS REPORTED IN INCLUDED REFERENCES

| Group authors and SDR | Mitigation technique | SDR architecture | Real-time? | Sampling rate* | Live channels** | Acceleration factors |
|---|---|---|---|---|---|---|
| Case study receiver [53], [54] | **Signal processing**: MMSE blind detector [12], [68] | **Category (4)**: Host PC, USB FE, C++ DLL & LV platform. | Yes | Up to 25 Msps | Up to 12 live channels** [68] | SIMD, LV parallelizable loops, multi-threading, etc. |
| Jafarnia-Jahromi et al. *GNSRx* [34] | **Antennas**: space-time correlation [79], arrays [83], [84], **Signal processing**: vector tracking [65], power analysis [62] | **Category (3)**: 3 channel NI FE, C++ modular design | N/A | Up to 25 Msps | Up to 22 total signals** | SIMD, multi-threading, GPU accelerators |
| Humphreys et al. *GRID* [29], GRID + spoofer/receiver, [71] | **Cryptography**: SCER [70], [72], **Correlation**: C/A and P(Y) [73]-[76], **Antennas**: AOA [80], [81] | **Category (2)**: GPP DSP in C++-based | Yes [72], [73], [76], [80] | 5.7 Msps | Up to 14 live channels** [72] | DSP accelerators, C++ bit-wise parallel operations, LUTs |
| Gao et al. *PyGNSS* [51] | **Signal processing**: MLE based on DP [66], and DTE [67], **Correlation**: multi-layer multi-receiver [77] | **Category (4)**: Python-based SDR. USRP N210 and GN3S [47] for offline recordings | N/A | Up to 5 Msps | N/A | N/A |
| Fernandez-Prades et al. *GNSS-SDR* [52] | **Signal processing**: SPREE [63], **Antennas**: [85]-[87] | **Category (4)**: Host PC, *GNU Radio* [60], Linux-based, C++ open source | Yes [63], [87] | Up to 10 Msps [63]. | One** [87] | Multi-threading, SIMD, modularity, C++ optimized libraries |

*Maximum reported sampling rate per SDR with interference mitigation solution.
**Includes real-time conventional and spoofed channels.

In [80], [81], authors successfully tested and implemented a dual-antenna receiver using *GRID*, which detects phase and AOA of authentic and potential spoofing signals. By knowing delta phases calculated using the known antenna location and their separation, the technique has shown the ability to discern between authentic and counterfeit signals.

In [73]-[76] authors used *GRID* in a different mitigation approach. They implemented an additional receiver, which observes the military GPS P(Y) signals for possible detection of attacks, assuming this additional receiver signal from P(Y) is clean. This detection algorithm exploits the known phase-quadrature relationship of the encrypted P(Y) code relative to the C/A code and other properties. Only [73] and [76] were able to run in real-time operation based on optimizations that were implemented in the receiver.

Finally, in [70], [72], authors use cryptographic-based techniques for spoofing detection and mitigation. Assuming navigation data has a cryptographic security code associated with it, it is possible for the spoofer to execute a SCER attack, in which the spoofer attempts to estimate the security code chip value for each GPS signal it intends to attack. The defending receiver attempts to detect possible SCER attacks by means of hypothesis tests related to noise levels and the received signal power. The augmented *GRID* receiver performed close to real-time for these detection techniques since a 2 ms delay was found on the SCER attack in [70], although [72] did report real-time. TEXBAT [78] database was used in all non-live experiments for this group of authors.

*3) Gao et al. (PyGNSS)*

Authors in [51] developed an advanced receiver based on vector tracking of concurrent GPS channels for added robustness. The concept is called multi-receiver vector tracking (MRVT), which can be seen as multiple GPS receivers jointly tracked through shared receiver states and a single navigational filter. The navigation filter is commonly used as a modified, jointly-shared, extended Kalman filter, to exploit joint tracking techniques. The receiver is non-real-time and is mainly implemented in Python, an open-source interpreted language that lacks optimizations but is highly configurable. The receiver design is highly modular for fast adaptations and prototyping.

As the states of each receiver are shared using MRVT, many receiver optimizations can be achieved. For example, MLE can be used for joint observations of clock bias and clock drift in the navigation solution. In [66], a DP technique based on MLE of raw observations on the MRVT was implemented. In [67], a DTE technique similar to the previous DP solution was also implemented using joint observations for phasor measurement units (PMUs) in power grids. These PMUs share a known location. Thus, fewer unknowns can be exploited for DTE techniques. In [77], a multi-layered multi-receiver architecture was also proposed, where shared states from the MRVT are used to detect spoofing across channels. Several detection tests were implemented based on power analysis of joint elements between the multi-receiver architecture.

*4) Fernandez-Prades (GNSS-SDR)*

*GNSS-SDR* [52] is an open-source real-time GNSS multi-frequency receiver. The receiver uses the *GNU Radio* [60] framework as its core implementation. The receiver is highly configurable, but at the same time has many software dependencies (including *GNU Radio*). It is based on a Linux environment and can utilize many FE options, such as USRP units from Ettus [88], GN3S sampler [47], NSL Primo [89], and IFEN's NavPort [90]. It can also work with recorded binary files. The architecture is based on modularity and MT, meaning each channel has its own independent acquisition, tracking and navigation thread. Once the receiver has enough channels, it can then compute PVT solutions. The receiver's performance depends on the host PC's computational capabilities, but still

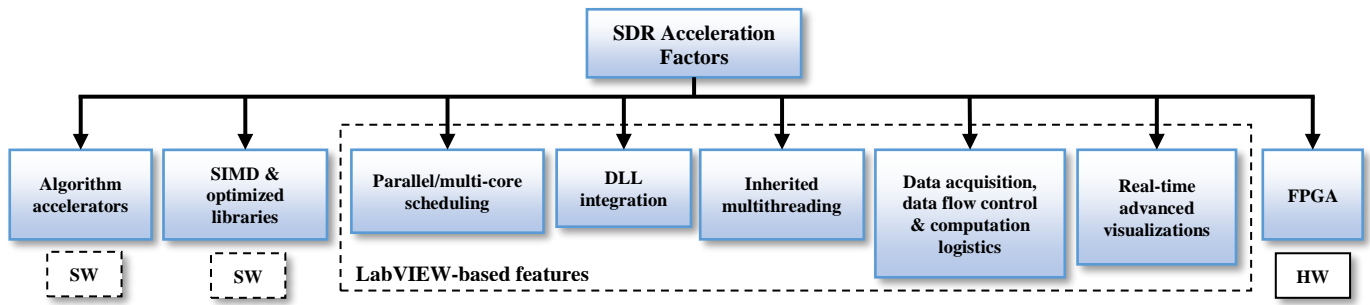

Fig. 2. Software-defined GPS receiver platform accelerator factors for real-time operation.

leverages several acceleration factors such as SIMD and vector-based operations in Linux [52].

Due to their robustness, there has been considerable research in the use of antenna arrays for interference detection [85]-[87]. In these works, the authors built and implemented an 8-channel antenna array, and tested it using several scenarios, both during real-time operation using the *GNSS-SDR* receiver and a commercial off-the-shelf (COTS) GPS simulator, as well as with offline recordings [87]. They were able to run algorithms with enhanced acquisition and detection of spoofers using an antenna array, in real-time on a single tracking channel.

Since the receiver is open-source, authors in [63] implemented real-time interference detection and mitigation on the same software receiver using a signal processing technique. They added a parallel tracking channel for the same PRN code to detect a second correlation peak that might be deviating from the main peak, thus, detecting a spoofer attempting to avert the receiver to an erroneous PVT solution. In this work, authors used TEXBAT [78] recordings for offline testing as well as real-time operation with *GNSS-SDR*.

## IV. A CASE STUDY RECEIVER BUDGETING WITHOUT INTERFERENCE

Before introducing advanced mitigation techniques, a receiver used for the case study is explored. A LV-based GPS L1 single-frequency receiver reported by authors in [53], [54]. It implements baseband functionality in C/C++ units that have been compiled as dynamic link libraries (DLLs). The *P-SW platform* architecture uses several LV-based acceleration factors such as parallelizable loops, inherent MT, and other non-LV-based enhancements, such as SIMD *Intrinsics* from Intel [91] to exploit the host CPU, as seen in Fig. 2. For algorithm accelerators, an advanced acquisition module was implemented in SW based on [57]. For optimized libraries, *FFTW* [92] and *Eigen* [93] packages are used for FFT and matrix operations, respectively. LV allows FE interfacing and easy control based on automatic parallelization blocks that allow real-time operation of the receiver. Advanced visualizations and GUI controls were also used for real-time debugging and monitoring. Additional LV-based accelerations can be achieved via FPGA. As seen in Table III, up to 25 Msps can be achieved, and up to 12 live tracking and/or spoofing channels are attained.

For the case study interference mitigation algorithm implemented, a *common configuration* in the receiver is used: sampling rate at 5 MHz, *int8* data type, and I-Q interleaved samples. For further details on SW and HW components, as well as performance tests, the reader is directed to [53], [54].

In the following, a *computational complexity budget* breakdown of common operations is examined in the case study receiver. It will be followed in Section VI by a computational complexity budget analysis in interference mitigation mode to demonstrate a noticeable surge in computational load.

To assess instrumentation capabilities of the case study receiver, several offline computations are benchmarked to obtain the aforementioned budget of common GPS baseband operations. These operations, although in offline mode, provide a very close benchmark value to real-time execution. In the context of host PC memory allocations and data type handling, previously mentioned common configurations are used for all budget operations. These common configuration parameters become relevant in terms of (linear) scalability of basic arithmetic operations, as well as memory allocation.

Sample data type is also important, as it specifies the number of quantization bits for sample resolution [47], [58]. In GPS SDR solutions, 8-bit samples have been proven to be enough for nominal operation and good precision. Some solutions use even fewer bits for faster computations [41], but since *P-SW platform* SDRs use multi-purpose processors, the byte (8-bit) fits naturally.

As an example, using said common configuration, a 1-ms block of data would consist of 10,000 bytes since each sample is of byte size (int8), and 1 ms of raw data collected at 5 MHz sampling rate consists of 5,000 samples for both the in-phase and quadrature interleaved components. Many FEs use intermediate frequency (IF) mixers, which output IF in-phase only samples, thus, receiving a smaller block of samples but also requiring an extra down conversion step to get the baseband I-Q sample pair. For the case study SDR, the FE uses a direct down-converter, thus, providing samples in baseband as an I-Q pair to save on extra operations. In terms of FFT operations, using a next power-of-two FFT size is common practice in SDR receivers. For 5 MHz sampling rate, the next FFT size would be of 8,192 for a single 1-ms epoch. For a sampling rate increase to 10 MHz, the numbers would scale up by a factor of two, and a 1-ms block would now be of size 20,000 samples, considering previous logic. This also means the correlators would need to deal with twice the number of operations, depending on their implementation. The FFT size would also jump to 16,384. The number of these operations provide an idea of computational budget for common SDR operations in real-time operation.

Table IV shows a computational budget for the case study SDR. A full acquisition of a single PRN search benchmark is included. As mentioned previously, the acquisition operation executes on 4-ms integration time and a 10 KHz frequency search band, respectively. More on FFT sizes and the algorithm implemented can be seen in [46], [57]. For common tracking operations, conventional correlators were benchmarked, i.e., in-phase and quadrature early, prompt, and late. This totals 6 correlators per GPS channel. As for the discriminators, a 2nd order delay-lock loop and phase-lock loop, and 1st order frequency-lock loop were used, respectively [58]. Therefore, a full tracking cycle contains carrier generation and wipe-off, 6 correlators processing and integration, and common computations, i.e., *atan* for discriminators. For a single correlator, all these previous operations were considered after the carrier wipe-off. Navigation benchmarks are not included since they account for minimal expenses as PVT solutions run at 2 Hz.

## V. Chosen Interference Mitigation Algorithm For Receiver Budgeting

This section expands on a modified interference mitigation algorithm [68], which is a computationally optimized version of [12]. There are many other spoofing methods, nevertheless and without the loss of generality, the following algorithm is chosen because of the noticeable computational surge (times) of receiver operation in mitigation mode.

### A. Background

The overall function of a GPS receiver is to synchronize to satellites to obtain navigation data and extract measurements for range estimations. The tracking synchronization is performed by correlating received (spreading) signals with locally generated (despreading) replicas of expected satellite PRN codes to maintain their alignment. A GPS received signal is composed of a superposition of multiple satellite codes contaminated by channel distortions and spoofing interference.

Conventional GPS receivers employ replica codes, which are the same as those sent by satellite transmitters. This ensures acceptable reception in outdoor areas but does not provide sufficient immunity against spoofing. There exist advanced solutions that enhance local (despreading) codes in the receivers to reduce these effects. De-correlators and MMSE detectors [94] are typically used for mitigating the multiple access interference effects in spread spectrum systems. However, the MMSE approach is computationally intensive due to required autocorrelation matrix inversions. To simplify the processing, some investigators estimate the cross-correlated signal and subtract it from the weak signal channel [58], [95], [96]. This solution is not optimal, computational overhead is still high, and applies only to known jamming signals. An ad-hoc solution is suggested in [97] using an additional orthogonalization process for better separation of weak and strong channels. There, the authors observed the performance of their method deteriorates for various conditions of available stronger jammers. An optimal solution is presented in [12], where the authors propose an algorithm for GPS-like interferer mitigation. This method is further optimized computationally

TABLE IV
COMPUTATIONAL COMPLEXITY BUDGET FOR CONVENTIONAL SDR
OPERATIONS AT 5 MHZ SAMPLING RATE

| Operation Type | Cycle time (ns) |
|---|---|
| FFT 8,192 (1 ms) | 46,547.2 |
| FFT 32,768 (4 ms) | 399,186.2 |
| Full acquisition (single PRN) | 5,684,375.5 |
| Single correlator* | 1,689.03 |
| Full tracking epoch** | 18,635.0 |

*Single correlator includes code generation, mixing, and integration cycle time, for, e.g., early in-phase.
**Full tracking includes carrier wipe-off, 6 conventional correlators for in-phase and quadrature, for early, prompt and late, respectively, final integration, and discriminator output.

next and used as a case study for budgeting the operation complexity of the spoofer-mitigating SDR receiver.

### B. Mitigation algorithm description

In the following, the signal model described in [12] is explored. Without loss of generality, it is assumed the GPS receiver is already coarsely synchronized with available GPS signals using an acquisition process, and each satellite signal is being tracked by finely aligning the received signal with a corresponding locally-generated replica, as conventionally applied in GPS receivers in nominal conditions.

Each satellite is assigned a dedicated processing channel $k \in \{1,...,K\}$. The received sampled signal for each channel, $k$, after carrier wipe-off and prior to code correlation is denoted as vector $\mathbf{r}_k$ of length N samples for one code period. This signal is a linear additive combination of the synchronized $k$-th PRN satellite signal $\mathbf{s}_k^0$ of one code period length, power $p_k$, and sinusoidal and code modulated signals from other $K-1$ visible satellites, plus noise. Additionally, the $k$-th satellite signal is modulated by a navigation bit-sample, $b_k$, such that:

$$\mathbf{r}_k = b_k \sqrt{p_k} \mathbf{s}_k^0 + \mathbf{i}_k + \mathbf{n}_k \quad (1)$$

where $\mathbf{i}_k$ is the interference of other $K-1$ satellite signals and $\mathbf{n}_k$ is the noise.

The receiver wipes-offs (despreads) the codes by multiplying to a despreading code, $\mathbf{h}_k$, and integrating (i.e., a correlator is used to associate received signal and dispreading replica). The decision variable $d_k$ for the channel $k$ is then determined to be the following:

$$d_k = \mathbf{h}_k^T \mathbf{r}_k \quad (2)$$

where $\mathbf{h}_k$ is a unit norm vector that does not amplify or attenuate the received power during a bit-sample. Each receiver channel minimizes its mean squared error (MSE) cost function, denoted as $MSE_k$ to determine an optimal dispreading code [12]. The MSE for each satellite $k$ is as follows:

$$MSE_k = E\left[\left(b_k - \frac{d_k}{\sqrt{p_k}}\right)^2\right] \quad (3)$$

This can be interpreted as a normalized MSE in comparison with the definition used in [98]. Applying Karush-Kuhn-Tucker (KKT) conditions, the resulting optimal despreading code solution is given as [12]:

$$\mathbf{h}_k = p_k \mathbf{R}_k^{-1} \mathbf{s}_k^0 \quad (4)$$

where $\mathbf{R}_k = E[\mathbf{r}_k \mathbf{r}_k^T]$ is the autocorrelation matrix of the received signal of channel $k$ after carrier synchronization, and $\mathbf{R}_k^{-1}$ is its inverse or pseudoinverse for minimum norm solution if the matrix is singular.

In [12], authors proposed a group-weighting method that could trade off complexity vs. performance in the code adaptation. The next subsection describes a computationally optimized version of this technique, as seen in [68].

### C. Reduced complexity despreading approach

We restrict the decoding sequence to the following format:
$$\mathbf{h}_k = \mathbf{w}_k (.*) \mathbf{s}_k^0 \quad (5)$$

$$\mathbf{w}_k = \left[ (w_{1,1},\ldots,w_{1,g}), (w_{2,1},\ldots,w_{2,g}), \ldots, (w_{M,1},\ldots,w_{M,g}) \right]^T \quad (6)$$

where $(.*)$ is the element-by-element multiplication of the vectors, and $w_{i,j}$ is the $j^{th}$ element of the $i^{th}$ group of size $g$, making $\mathbf{w}_k$ a vector of size $M = \dfrac{N}{g}$. As the elements of $\mathbf{s}_k^0$ are $\pm 1$, then $\mathbf{h}_k^T \mathbf{s}_k^0 = g \mathbf{w}_k^T \mathbf{1}_M$. Let us split $\mathbf{h}_k$, $\mathbf{r}_k$ and $\mathbf{s}_k^0$ into $M$ segments:

$$\mathbf{h}_k = \begin{bmatrix} \mathbf{h}_{k1} \\ \mathbf{h}_{k2} \\ \vdots \\ \mathbf{h}_{kM} \end{bmatrix}; \mathbf{r}_k = \begin{bmatrix} \mathbf{r}_{k1} \\ \mathbf{r}_{k2} \\ \vdots \\ \mathbf{r}_{kM} \end{bmatrix}; \mathbf{s}_k^0 = \begin{bmatrix} \mathbf{s}_{k1}^0 \\ \mathbf{s}_{k2}^0 \\ \vdots \\ \mathbf{s}_{kM}^0 \end{bmatrix} \quad (7)$$

The constrained $\mathbf{h}_k$ will be then as follows:

$$\mathbf{h}_k = \begin{bmatrix} \mathbf{h}_{k1} \\ \mathbf{h}_{k2} \\ \vdots \\ \mathbf{h}_{kM} \end{bmatrix} = \begin{bmatrix} w_1 \mathbf{s}_{k1}^0 \\ w_2 \mathbf{s}_{k2}^0 \\ \vdots \\ w_M \mathbf{s}_{kM}^0 \end{bmatrix} \quad (8)$$

Denote $c_{kj} = \mathbf{h}_{kj}^T \mathbf{r}_{kj}$ as a partial correlation, and $\mathbf{c}_k = [c_{k1}, c_{k2}, \ldots, c_{kM}]^T$ as a vector of partial correlations. Then,
$$d_k = \mathbf{h}_k^T \mathbf{r}_k = \mathbf{w}_k^T \mathbf{c}_k, \text{ and} \quad (9)$$

$$MSE_k = 1 - 2g \mathbf{w}_k^T \mathbf{1}_M + E\left[ \dfrac{(\mathbf{w}_k^T \mathbf{c}_k)(\mathbf{c}_k^T \mathbf{w}_k)}{p_k} \right]$$
$$= 1 - 2g \mathbf{w}_k^T \mathbf{1}_M + \dfrac{1}{p_k} \mathbf{w}_k^T E\left[ \mathbf{c}_k \mathbf{c}_k^T \right] \mathbf{w}_k \quad (10)$$
$$= 1 - 2g \mathbf{w}_k^T \mathbf{1}_M + \dfrac{\mathbf{w}_k^T \mathbf{R}_{ck} \mathbf{w}_k}{p_k}$$

where $\mathbf{R}_{ck}$ is the autocorrelation of partial correlations. Minimization of $MSE_k$ is achieved by applying KKT conditions, taking the partial derivative with respect to $\mathbf{w}_k$, and equating to zero:

$$\dfrac{2 \mathbf{R}_{ck} \mathbf{w}_k}{p_k} - 2g \mathbf{w}_k^T \mathbf{1}_M = 0 \quad (11)$$

$$\mathbf{w}_k = g p_k \mathbf{R}_{ck}^{-1} \mathbf{1}_M \quad (12)$$

These solutions are suboptimal for $g > 1$, in comparison with those provided by the optimal algorithm from (4). This is a result of the group-weighting method decreasing the freedom of designing dispreading code; alternatively, it has the advantage of significantly decreasing the computational complexity. Moreover, the solution presented for partial correlations can also be implemented with a computational complexity of $O(M^3)$. This solution has a computational gain, of order $O(g^3)$, in comparison with the conventional MMSE solution from (4). Fig. 3 shows the interaction of partial correlations $\mathbf{c}_k$ with input and replica samples, generating the optimal output sample $d_k$. Ultimately, these $M$ weights, $\mathbf{w}_k$, are correlated to obtain the maximum SINR solution in decision variable $d_k$, and are implemented in the tracking correlators for ideal navigation bit extraction.

The quality of a GNSS receiver operation is defined by the accuracy of the PVT solution. The presented algorithm addresses *pre-PVT* detection and mitigation of the interference and falls in the signal processing-based category. It applies a scalable blind equalizer-like technique to sense and remove certain interference *per channel*. The performance of this is not based on bit recovery but on bit-error rate, as is used to assess its quality.

## VI. CASE STUDY RECEIVER BUDGETING WITH INTERFERENCE

This chapter describes the implementation of the MMSE algorithm from Section V as an advanced correlator unit in the case study receiver. It further elaborates on the implementation as seen in [68] and addresses complexity through diverse optimization techniques. It describes a computational budget with said interference mitigation algorithm to complement the discussion from Section IV.

The MMSE correlator is presented as a modification to the conventional GPS tracking correlator. For this implementation, the in-phase prompt (IP) channel is adjusted. An essential aspect of the MMSE algorithm is the computation of an autocorrelation matrix seen in (12), which is obtained from a single correlator output. The autocorrelation matrix, obtained from the outputs of the partial correlator, serves as a *blind sensing* module for possible interferences. It senses cross-correlation anomalies in the matrix and adjusts weights accordingly for interference filtering. Some options to control the computational complexity of these updates are the grouping parameter $g$ and window size $L$, which are detailed next.

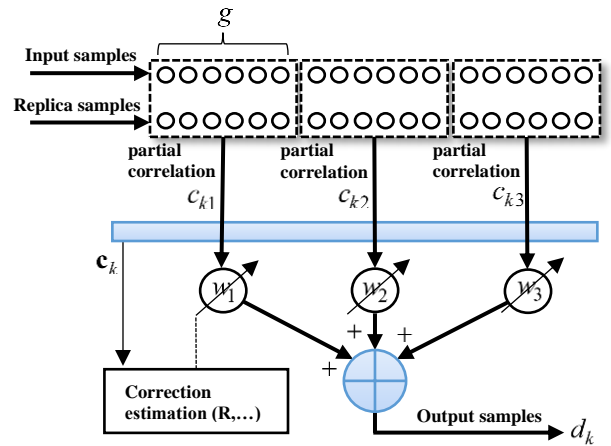

Fig. 3. Group-weighting method partial correlation sample interaction [68].

### A. MMSE correlator implementation

Fig. 4 shows the MMSE correlator as an advanced unit, which can be added to the existing tracking loops. Specifically, it modifies a traditional correlator unit so there are partial integrate and dump (I&D) and complete I&D filters instead of common I&D filters to extract samples between said filters. This modified MMSE correlator is used to compute, after carrier wipe-off, the autocorrelation matrix of partial correlations, $\mathbf{R}_{ck}$, corresponding to the method in (12). Using optimal weights, the MMSE correlator outputs an integrated value with filtered-out cross-correlation interference.

Conventional tracking loops may have variable signal integration lengths (in samples) which depend on Doppler shifts. In the case study SDR, the tracking loop maintains a constant length of 1023 chips, using a pre-integration of samples within one chip, to address Doppler shifts, e.g., one may have 4 or 5 samples per chip at a sampling rate of 5 MHz depending on the Doppler effect. Initially, the tracking loop collects a fixed size block of data corresponding to 1 ms. Thus, the locally-generated carrier is also sampled at 5 MHz, which is equivalent to approximately 5000 samples per epoch, again including Doppler effects. When the carrier is wiped off, the samples should be pre-integrated within each chip, thus, ending with an array of $N = 1023$ chips. For the group-weighting method, the array is up-sampled to $N = 1024$ so the size is a power-of-two, and the chips can be grouped using $g$, as will be seen in Section V-C.

After chip pre-integration and up-sampling, the received vector, $\mathbf{r}_k$, is mixed with aligned in-prompt code replica, $\mathbf{s}_k^0$, for code alignment. The next step involves a partial integrate and dump block, which integrates a sequence of chips into $M$ partial correlation groups, outputting the $1 \times M$ $\mathbf{c}_k$ vector as seen in Fig. 4.

The $\mathbf{R}_{ck}$ *accum.* block collects the $\mathbf{c}_k$ vector for statistical matrix estimation on every 1-ms epoch. Once the autocorrelation matrix, $\mathbf{R}_{ck}$, has collected enough epoch-vectors, thus, reaching a suitable statistical significance, the MMSE solution can be linearly computed using (12) for the optimal group-weighting solution. This is done at the MMSE algorithm block. Once the matrix is suitable for inversion, it computes the solution on every epoch iteration. This block then outputs optimal $\mathbf{w}_k$ coefficients needed for interference filtering based on potential cross-correlation interference patterns seen in the partial autocorrelation matrix, $\mathbf{R}_{ck}$. The $\mathbf{w}_k$ coefficients are of size $M$, so they are matched to the received signal partial correlation vector, $\mathbf{c}_k$, for corrections. Other inputs to the MMSE block are group size, $g$, and channel

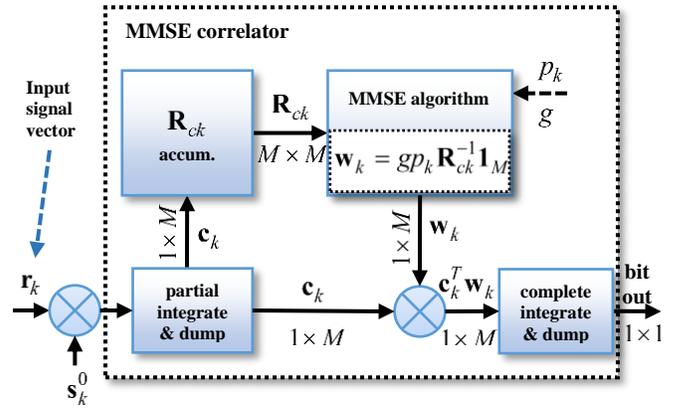

Fig. 4. A single MMSE correlator implementation in SDR tracking loop after carrier wipe-off coming from the IP channel [68].

signal power, $p_k$, the latter of which is estimated continuously on conventional tracking loops.

Last, another complete integrate and dump block is used to integrate the result $\mathbf{w}_k \cdot \mathbf{c}_k$ for an optimal bit-sample MMSE correlator output. Since IP channel is used, this output corresponds to navigation data.

As mentioned, three pre-integration steps are involved in the modified MMSE correlator: (1) chip pre-integration from samples to chips, (2) partial integration from $N$ to $M$, based on the grouping parameter $g$, and (3) complete integration from $M$ to 1 for the navigation bit-sample. The first two integrations are done in one step at the partial integrate and dump block shown in Fig. 4 for reduced complexity implementation. Also, when comparing said modifications to a conventional GPS correlator, the MMSE correlator can be seen as a single unit replacing classical GPS blocks. Additionally, this unit can be implemented either as a replacement to a conventional correlator or as an additional unit in the receiver.

### B. Recursive autocorrelation matrix computation

The implementation of the $\mathbf{R}_{ck}$ accum. block seen in Fig. 4 is achieved by a recursive statistical method with attainable computational complexity for the current SDR testbed. Fig. 5 shows a detailed implementation of the $\mathbf{R}_{ck}$ accum. block. Consider a sequence of partial correlation vectors $\mathbf{c}_k(t_l)$, $l = 1,...,t_L$, that are used to estimate autocorrelation matrix $\mathbf{R}_{c_k}$, given as:

$$\mathbf{R}_{ck}(t_L) = \frac{1}{L}\sum_{l=1}^{L}\mathbf{c}_k(t_l)\mathbf{c}_k(t_l)^T \quad (13)$$

where $t_l$ is the time instant of $\mathbf{c}_k(t_l)$ epoch-vector availability, and $L$ is the window size, which is fixed during receiver initialization. To address complexity constraints for the case study SDR implementation, the autocorrelation matrix $\mathbf{R}_{ck}$, seen in (12), is computed recursively in a sliding window manner by discarding the oldest $\mathbf{c}_k(t_l)$ entry and adding the newest one as follows:

$$\mathbf{R}_{ck}(t_L+1) = \mathbf{R}_{ck}(t_L)$$
$$-\frac{1}{L}\mathbf{c}_k(t_1)\mathbf{c}_k(t_1)^T \quad (14)$$
$$+\frac{1}{L}\mathbf{c}_k(t_L+1)\mathbf{c}_k(t_L+1)^T$$

This recursively computed matrix is based on a sliding window technique explained as follows. The $\mathbf{R}_{ck}$ accum. block seen in Fig. 4 receives as input vectors of partial correlations, $\mathbf{c}_k$, of length $M$ which are fed into a *first-input first-output (FIFO) vector buffer* of size $L$ vectors, and $M \times L$ samples. On each epoch, a new input vector of partial correlations, $\mathbf{c}_k(t_{new})$, is received for a channel $k$, which is fed into the FIFO vector buffer. At the same time, the oldest value on the FIFO vector buffer is pushed out, thus keeping the buffer at a fixed window size, $L$. At the same time, the input vector of partial correlations, $\mathbf{c}_k(t_{new})$, is multiplied by its transpose to generate a matrix of size $M \times M$ to be added to the $\mathbf{R}_{ck}$ accumulator seen in Fig. 5. Similarly, the oldest output vector of the buffer, $\mathbf{c}_k(t_{old})$, is similarly subtracted from $\mathbf{R}_{ck}$ and then discarded. These additions and subtraction on the $\mathbf{R}_{ck}$ accum. block occur continuously on every 1-ms integration epoch, as per (14), until the FIFO vector buffer is full. This signifies a statistically sound autocorrelation matrix $\mathbf{R}_{ck}$.

From the C/C++ implementation perspective, both the size of the FIFO vector buffer as well as the $\mathbf{R}_{ck}$ matrix become relevant in terms of memory usage. The FIFO vector buffer and the matrix size are determined by the grouping parameter $g$, the vector length $M$, and the sliding window size $L$. The FIFO vector buffer block and $\mathbf{R}_{ck}$ matrix are both implemented using dynamic memory allocation on variables that are initialized when called from the LV DLL call library function blocks. These memory allocations depend on initial user parameters, $g$ and $L$, prior to runtime execution and are not adjustable during runtime execution. To adjust them, the receiver would have to be stopped, reconfigured, and reinitialized. There are calibrated recommendations for the grouping size $g$ and sliding window $L$ pairs: $g = 1$, $L = 1500$; $g = 2$, $L = 1000$; $g = 4$, $L = 1000$; $g = 8$, $L = 1000$; $g = 16$, $L = 500$; $g = 32$, $L = 100$; $g = 64$, $L = 300$. In this work, the configuration that achieves the best trade-off between computation complexity and performance is $g = 64$ and $L = 300$, which allocates a dynamic variable array of 4,800 samples. Another recommended configuration pair, but slightly heavier in computation complexity is $g = 64$ and $L = 1200$. This last pair consists of four times the window size for the same $g$ parameter and provides improved blind adaptation response, as the matrix becomes more robust and more sound in statistical terms, but at the cost of memory allocation. This configuration pair will be used as a case study in Section VII-B.

Since testing is done on navigation data mitigation abilities, a parallel MMSE correlator is used to bypass spoofing effects on synchronization loops. Therefore, this MMSE correlator unit is separated from nominal receiver operations to explore its instrumentation and mitigation capabilities at the correlator level and without disturbing common tracking operations and synchronization loops.

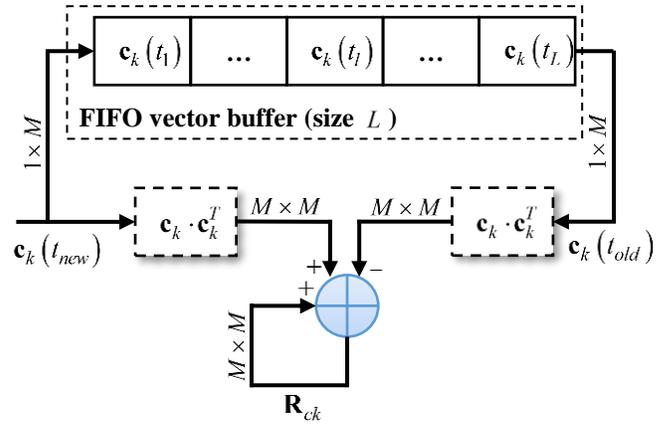

Fig. 5. Sliding window technique for recursive autocorrelation matrix computation and FIFO vector machine.

*C. Autocorrelation matrix numerical estimation*

The recursive autocorrelation matrix accumulated on every epoch is assumed to be full-rank and real valued. In realistic scenarios, however, it can be affected by numerical estimation errors and become ill-conditioned or singular. This paper follows [99], which applies a full-rank regularization technique based on eigenvalue spread. This regularization technique is a constrained minimum output energy estimator that substitutes $\mathbf{R}_{ck}^{-1}$ in (12) as follows:

$$\mathbf{w}_{CMOE,k} = gp_k \left(\mathbf{R}_{ck} + \nu \mathbf{I}_{MM}\right)^{-1} \mathbf{1}_M \quad (15)$$

where $\nu$ is a positive-valued aid parameter that moderates eigenvalue spread in the autocorrelation matrix, which is common in multi-user detection. This approach was tailored for the MMSE correlator unit by fixing a calibrated value as the aid parameter for invertibility. This technique was implemented internally in the receiver and is used continuously for every epoch computation of the solution in (12).

*D. MMSE correlator block operations and numerical approach*

It is easy to surpass computational capacity when linear algebra floating-point operations are involved, especially if real-time operation is required. The operation count for the individual blocks on the MMSE correlator unit (see Fig. 4), including the recursive autocorrelation matrix accumulator (see Fig. 5) and the reduced complexity solution (12) computation, is presented in floating-point operations (FLOPs) [100] in Table V. Without the loss of generality, we consider $N = 1024$ and a grouping parameter of $g = 64$ with a window size of $l = 300$, which proves to be a feasible approach on the case study receiver. The resulting size of the matrix is $M = 16 \times 16$. Table V shows an estimated count of FLOPs for each stage involved in the MMSE correlator unit. These operations are for every 1 ms (epoch) of a bit-sample. For the first step, which involves interference injection, up to three interferers can be configured, each one accounting for $N$ FLOPs. These interferers inject chip sequences of size $N$ with cross-correlation jamming effects. More on interference injection is discussed in Section

TABLE V
OPERATION COUNT (FLOPs) FOR SEVERAL STAGES ON MMSE CORRELATOR ON EVERY EPOCH

| Stage | Operation Count (FLOPs) |
|---|---|
| Interference Injection $\mathbf{i}_k$ | $N$ |
| Partial integrate and dump ($N \to M$) | $M(g-1)$ |
| Old $\mathbf{c}_k \mathbf{c}_k^T$ | $M^2$ |
| New $\mathbf{c}_k \mathbf{c}_k^T$ | $M^2$ |
| Accumulate $\mathbf{R}_{ck}$ | $3M^2$ |
| Compute (12) solution | $2M + M^2 + M^3$ |
| Integrate bit-sample $\mathbf{c}_k^T \mathbf{w}_k$ | $2M$ |

TABLE VI
REAL-TIME COMPUTATIONAL COMPLEXITY BUDGET IN TIME CYCLE OF 1 MS EPOCH

| Operation type | Scope (*corr* or *chan***) | Cycle time (ns) | Share of computational time* (%) |
|---|---|---|---|
| SIMD carrier generation and wipe-off | *chan* | 8,500.8 | 0.85% |
| SIMD code generation, wipe-off, and integration | *corr* | 1,655.7 | 0.17% |
| Tracking discriminators | *chan* | 200.0 | 0.02% |
| SIMD jamming injection | *corr* | 5,263.8 | 0.53% |
| MMSE pre-integrator block $\mathbf{c}_k$ *** | *corr* | 28,683.8 | 2.87% |
| Accumulate $\mathbf{R}_{ck}$ (14) | *corr* | 26,348.0 | 2.63% |
| MMSE solution (12) | *corr* | 8,502.2 | 0.85% |
| Integrate bit $\mathbf{c}_k^T \mathbf{w}_k$ | *corr* | 78.8 | 0.01% |

*Computing time period cycle of 100% is equivalent to 1,000,000 nsec or 1 ms.
**Scope of the operation is either *corr* (correlator channel) or *chan* (full GPS channel).
**Pre-integrator modifies received $\mathbf{r}_k$ vector of size, e.g., 5,000 samples at 5 MHz, to $\mathbf{c}_k$ vector of size 16.

VII-A. The partial correlations stage involves reducing the received signal vector ($N \to M$). The matrix accumulation (see Fig. 5) can be seen of order $O(M^2)$. The MMSE solution in (12) is an estimated approach on FLOPs since it contains a matrix inverse operation that requires to be computed on every epoch iteration. This stage accounts for the most operations as it is of order $O(M^3)$.

*E. MMSE unit integration budget*

Similar to conventional GPS computational complexity budget operations listed in Section IV, this section provides a budget including case study mitigation operations to assess complexity load. It also discretizes common real-time operations as individual units for comparison.

Tracking correlators and discriminators are commonly implemented in HW for faster operation since they are updating their discriminators and integrators on every given epoch (1 ms). Other advanced correlators have longer integration times, e.g., 10-20 ms, but are considered for advanced receivers. Thus, this 1 ms measurement becomes the threshold for real-time operation in SW. Therefore, all relevant computations occurring at each epoch including carrier and code generators, correlators, and tracking discriminators, should be computed in less than 1 ms for the SW receiver to be considered real-time. If this ensemble of conventional GPS computations can be processed in less than 1 ms of integration time, it leaves additional computational resources for aggregate operations, such as interference injection and mitigation algorithms similar to those seen in Table V.

Conventional tracking and MMSE correlator unit operations from Table V can be seen in Table VI as a real-time computational budget benchmarked from the case study receiver in mitigation mode. This computational budget relates all operation benchmarks to 1 ms, which represents 100% of the shared computations for the time period cycle.

SIMD-based correlators for carrier and code are shown for a single epoch using 1 ms integration time. For code correlation, a single unit benchmark is shown, i.e., IP correlator, as opposed to six correlators [58]. This encompasses code generation, wipe off, and integration. If the full nominal operation of a single GPS channel would be assessed, then 6 correlator values and a carrier value would need to be accounted for. Moreover, if 12 GPS channels are assumed to be tracked, then this number would be scaled properly. Tracking discriminator computations are also shown, which comprises feedback updates from tracking loops as discussed in Section IV. For interference injection, a SIMD jammer benchmark was added per correlator channel, representing a single interferer's computational expense. The pre-integrator of vector $\mathbf{c}_k$ is measured per correlator channel, and integrates from samples to pre-integrations, e.g., 5000 to 16, as mentioned in Section VI-A. The accumulation on $\mathbf{R}_{ck}$, as seen in Table V, represents the full computation seen in equation (14) for a single correlator channel. The MMSE solution in (12) was similarly evaluated for a single correlator channel. Finally, the full integration of optimized coefficients with grouping vector $\mathbf{c}_k$ for the navigation bit-sample is measured.

Table VI shows case study receiver tracking operations. It can be seen that a single GPS channel accounts for roughly ~1.9% of the total computational budget. Since the receiver can run up to 12 channels, this would be scaled to ~22.4%, thus, leaving more than 70% of the computational budget in conventional mode. Additionally, previous testing on case study SDR showed the LV environment would provide an additional overhead of approximately ~3%, which covers the FE interface and communication, and data acquisition among other common operations [46]. This ~3% is included in the total budget computations to approximate live capabilities of the receiver in conventional mode.

Alternatively, a single channel being jammed with one interferer and mitigated spends roughly ~8.8%, a 6.9% rise from the non-mitigation mode. Therefore, full implementation of the conventional operation, jamming, and mitigation mode on 12 live channels reaches the 100% threshold for real-time operation. This shows a 5 times surge in computational complexity in the case study receiver when in mitigation mode for the same common configuration. For additional comparison purposes, the SDR was able to achieve a maximum of 8 channels at 10 MHz sampling rate, with similar full implementation in mitigation mode, when contrasting a maximum configuration of 25 MHz sampling rate with 12 live channels in non-mitigation mode. These comparisons demonstrate a computational surge even for a fully optimized mitigation algorithm (see Section V) [68]. The next section assesses the performance of the algorithm filtering capabilities in terms of bit-error rate (BER) for a single channel when interference is added to real GPS signals.

## VII. SIMULATION RESULTS

In this work, specific interference is used to represent an idealized perfectly synchronized GPS-like spoofer with malicious intentions that affect cross-correlation of despreading codes. The case study method does not specifically detect the interference but adjusts when such phenomena are observed (blind estimator). The interference injected mimics a worst-case cross-correlation scenario.

The interference injection occurs right before the code wipe off and after the carrier wipe off; therefore, the interference signal does not contain Doppler effects. This is assumed to be the worst-case scenario for an attack since a spoofer would have fully modulated power to the intended in-phase channel. This creates the most damage, thus, can be used as a reference scenario. This approach makes the receiver immune to GPS-like synchronized intervention.

The case study SDR with interference injection and mitigation is evaluated by using a hybrid approach of real signals coming from a LabVIEW-based NI GPS simulation toolkit [101] and internally generated synthetic interference. The method in [12] addresses a type of spoofer that can generate an exact same but time-shifted signal per channel using worst-case scenario delays, which have the greatest effect. Since code delays and Doppler shifts are known in the interference injection, a potential spoofer can use it to its advantage by filtering signals into the channel via cross-correlation interference.

Specifically, interference injection occurs by replicating the $k$-th channel code replica, $\mathbf{s}_k^0$, and delaying it a known number of chips. The delayed code replica, $\mathbf{s}_k^\alpha$ (where $\alpha$ is a delay in chips), is based on known cross-correlation properties of the PRN sequence for satellite $k$, thus, spoofing the modulated navigation bits coming from GPS satellites. The spoofing power amplifies when at the higher sidelobes of the gold code cross-correlation. This causes the correlation to lose orthogonality [102]. Therefore, the most efficient spoofing can occur at these sidelobes with delay $\alpha$.

The interferer navigation bits are modulated onto the spoofing signal by assigning random bit polarity every 20 bit-samples. The authentic and spoofed navigation bits are not necessarily synchronized in terms of bit edges since the spoofing bits can be injected at any time.

As mentioned in Section VI-E, the case study receiver can successfully jam up to 12 live channels. The grouping parameter was set to $g=64$ with a window size of $l=300$. As an aggregate experiment, we also tested $g=64$ and $l=1200$, which is four times the window size of previous parameters as discussed in Section VI-B.

### A. Performance results

In this subsection, we assess previous work [12] in a case study receiver for algorithm functionality and real-time operation [68] per channel for pre-PVT mitigation. Without the loss of generality, we simulate interference power relative to signal power in dB scale. Interference-to-signal ratio (ISR) noise power can be neglected when compared to the sum of the interference and signal powers after channel synchronization and integration.

We assess testing scenarios for matched filter (MF) correlators [58], and the case study MMSE correlator for a single channel. BER vs. ISR performance curves are plotted and evaluated. This is done to study the cleaning capabilities of the algorithm in the presence of strong interference, compared to a MF correlator. Similarly, a worst-case scenario of three jamming signals per channel, where all three jammers match in chip and bit alignment as well, is included.

Our goal is to obtain statistically significant BER performance results for functionality purposes in the chosen case study mitigation technique. Thus, we simulate 3 million navigation bits or 60 million bit-samples, corresponding to 1,000 min of simulation. We perform the simulations with a 200-second real signal file previously recorded from the NI GPS simulator [101]. The recorded file has strong GPS signals; thus, noise can be neglected as previously mentioned. We proceed to run the file 300 times to obtain our navigation bits quota while injecting interference bits with relative power levels from reported SNR from target channel, and random bit polarity. This is all done on the in-house LV-based SDR [53], [54], which was prototyped for this test as in [68].

A well-known cross-correlation *immunity* of around 26 dB is present in MF correlators based on GPS system design [1]. Nevertheless, after or around this ISR level (26 dB) the channel is completely corrupted. Fig. 6 shows the performance curve for one and three interfering signals, where BER has been assessed for 20 bit-samples corresponding to a full navigation bit. The curve compares the performance of MF vs. MMSE correlators. It also compares a MMSE correlator with 4 times the proposed window length discussed in Section VI-B. The comparison approach assesses the BER performance at a 30 dB ISR level (see vertical dashed line), which is well above the aforementioned 26-dB immunity level. This 30 dB ISR level would notably corrupt the GPS channel as it can be clearly seen for the cases of *MF (1)* and *MF (3)*, corresponding to MF with 1 interferer, and MF with 3 interferers present, respectively. For both these cases, the channel is corrupted with a BER of 50%.

When comparing against the MMSE correlator, we can see *MMSE (1)* shows a BER gain of almost $10^5$ against *MF (1)*, for 1 interferer. For *MMSE (3)*, a BER gain of around $10^{1.5}$ can be seen against *MF (3)*, for 3 interferers. Finally, when increasing the window size four times used in *MMSE (3)*, i.e., using *MMSE 4x (3)*, a BER gain of around $10^5$ is seen against *MF (3)*, similar to the gain seen in *MMSE (1)* vs. *MF (1)*, but for 3 interferers. Table VII summarizes these BER performance results at the 30 dB ISR observation dashed line (see Fig. 6).

## VIII. CONCLUSION AND FUTURE WORK

This paper provides an overall review and categorization of reported GNSS-SDR receivers for conventional and spoofing domains, in terms of reported instrumentation. Many conventional SDRs are not able to support mitigation modes in real-time, which necessitates separate categorization of spoofing mitigation receivers. A case study demonstration of a significant surge in complexity from a conventional to a mitigation receiver is presented. This is done through computational budgeting of an SDR receiver employing a modified MMSE algorithm from [12] optimized for real-time operation [68] in a LV-based receiver [53], [54]. The expanded version of [68] is presented, where the (pre-PVT) mitigation is accomplished through a blind equalizer processing. The approach was validated and assessed using BER performance curves, as [12] did not demonstrate operation with real signals and in real-time. It is selected due to demanding computational complexity. To analyze said complexity increase the paper tabulates a breakdown of common algorithmic operations for conventional and mitigation mode of the case study receiver. A surge of the computational load from 20% to 100% (5 times) of the case study receiver was seen when in mitigation mode as compared to conventional operation. A common configuration of 5 MHz sampling rate, int8 I-Q interleaved samples, and 12 live channels were used. It should be mentioned that the MMSE correlator unit could potentially be incorporated *in lieu* of any of the conventional correlators.

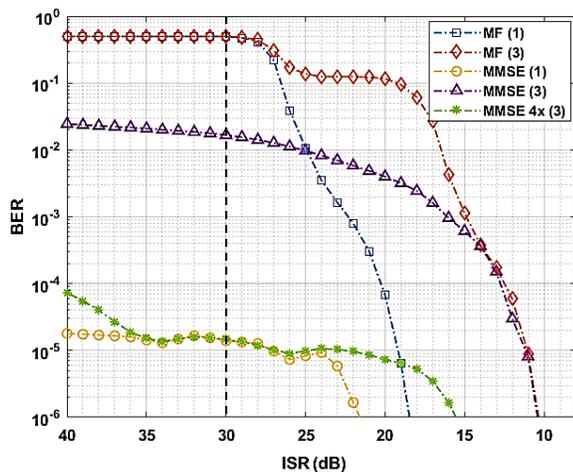

Fig. 6. BER vs ISR performance results for one and three interferers for MF and case study MMSE approach [68].

TABLE VII
RECEIVER BER PERFORMANCE FOR SPECIFIED TARGET ISR= 30 DB TO ASSESS GAINS AFTER GPS IMMUNITY LEVELS

|  | MF (1) | MMSE (1) | MF (3) | MMSE (3) | MMSE 4x (3) |
|---|---|---|---|---|---|
| ISR=30dB | $10^{0.3}$ | $10^{-4.9}$ | $10^{0.3}$ | $10^{-1.8}$ | $10^{-4.9}$ |

The case study employed an idealized spoofer perfectly synchronized in both carrier frequency and phase with the authentic signal, which allowed to concentrate on computational budgeting aspects and not on spoofing-specific effects. At the same time, the selected idealized environment can simulate worst-case scenarios as a baseline. Authors currently address spoofing mitigation aspects more comprehensively including sophisticated advancements of [12], [68], which will be presented in the future dedicated work. This projected work will align various methods with state-of-the-art spoofing mitigation techniques and test with representative signals such as TEXBAT [78].

In summary, when tightly integrating interference mitigation algorithms, especially in real-time manner, the complexity surge truly differentiates this class of GNSS-SDR receivers.

IX. ACKNOWLEDGEMENTS

The authors would like to thank anonymous reviewers for their valuable assessment in improving this paper.

REFERENCES

[1] P. Misra, P. Enge, Global Positioning System: Signals, Measurements, and Performance, Lincoln, Massachusetts: Ganga-Jamuna Press. 2012.
[2] C.J. Hegarty, E. Chatre, "Evolution of the Global Navigation Satellite Systems (GNSS)," *Proc. of IEEE*, Vol.96, No.12, pp.1902-1917, 2008.
[3] I. Skog and P. Händel, "In-Car Positioning and Navigation Technologies—A Survey," *IEEE Trans. Intell. Transp. Syst.*, vol. 10, no.1, pp. 4-21, 2009.
[4] K. J. Zou, K. W. Yang, M. Wang, B. Ren, J. Hu, J. Zhang, M. Hua, and X. You, "Network Synchronization for Dense Small Cell Networks," *IEEE Wireless Commun.*, vol.22, no.2, pp.108-117, 2015.
[5] P.A. Crossley, H. Guo, Z. Ma, "Time synchronization for transmission substations using GPS and IEEE 1588," *CSEE Journal of Power and Energy Systems*, vol.2, no.3, pp. 91-99, 2016.
[6] A. Carta, N. Locci, C. Muscas, and S. Sulis, "A flexible GPS-based system for synchronized phasor measurement in electric distribution networks," *IEEE Trans. Instrum. Meas.*, vol. 57, pp. 2450-2456, 2008.
[7] M. L. Psiaki, and T. E. Humphreys, "GNSS Spoofing and Detection," *Proc. IEEE*, vol. 104, no. 6, pp. 1258-1270, Jun. 2016.
[8] D. Schmidt, K. Radke, S. Camtepe, E. Foo, and M. Ren, "A Survey and Analysis of the GNSS Spoofing Threat and Countermeasures," *ACM Comput. Surv.*, vol. 48, no. 4, May 2016, doi: http://dx.doi.org/10.1145/2897166.
[9] M. G. Amin, and W. Sun, "A novel interference suppression scheme for global navigation satellite systems using antenna array," *IEEE J. Sel. Areas Commun.*, vol. 23, no. 5, pp. 999-1012, May 2005.
[10] R. L. Fante, and J. J. Vaccaro, "Wideband cancellation of interference in a GPS receive array," *IEEE Trans. Aerosp. Electron. Syst.*, vol. 36, no. 2, pp. 549-564, Apr. 2000.
[11] J. Arribas, C. Fernandez-Prades, and P. Closas, "Antenna array based GNSS signal acquisition for interference mitigation," *IEEE Trans. Aerosp. Electron. Syst.*, vol. 49, no. 1, pp. 223-243, Jan. 2013.
[12] C. Lacatus, D. Akopian, and M. Shadaram, "Reduced complexity crosscorrelation interference mitigation in GPS-enabled collaborative ad-hoc wireless networks - Theory," *Computers and Electrical Engineering*, vol. 38, no. 3, pp. 603-615, May 2012.
[13] G. Seco-Granados, J. A. Fernandez-Rubio, and C. Fernandez-Prades, "ML estimator and hybrid beamformer for multipath and interference mitigation in GNSS receivers," *IEEE Trans. Signal Process.*, vol. 53, no. 3, pp. 1194-1208, 2005.
[14] D. Borio, "GNSS acquisition in the presence of continuous wave interference," *IEEE Trans. Aerosp. Electron. Syst.*, vol. 46, no. 1, Jan. 2010.
[15] P. Closas, C. Fernandez-Prades, and J. A. Fernandez-Rubio, "Maximum Likelihood Estimation of Position in GNSS," *IEEE Signal Process. Lett.*, vol. 14, no. 5, pp. 359-362, May 2007.
[16] S. Savasta, L. L. Presti, and M. Rao, "Interference mitigation in GNSS receivers by a time-frequency approach," *IEEE Trans. Aerosp. Electron. Syst.*, vol. 49, no. 1, pp. 415-438, Jan. 2013.
[17] F. Dominici, P. Mulassano, D. Margaria, and K. Charqane, "SAT-SURF and SAT-SURFER novel hardware and software platform for research and education on satellite navigation," in *European Navigation Conference on Global Navigation Satellite Systems (ENC GNSS)*, Naples, Italy, May 2009.
[18] P. Fenton, B. Falkenberg, T. Ford, K. Ng, and A. J. V. Dierendonck, "Novatel's GPS receiver the high-performance OEM sensor of the future," in *Proceedings of the International Technical Meeting of the Satellite Division of the Institute of Navigation (ION GNSS '91)*, Albuquerque, NM, USA, 1991.
[19] W. Roberts, M. Bavaro, E. D. Tijero, F. Legrand, S. Vaccaro, A. Sage, and C. Hill, "PRECISIO-Design considerations for a multi-constellation,

- [19] multi-frequency software receiver," in *Satellite Navigation Technologies and European Workshop on GNSS Signals and Signal Processing (NAVITEC), 2010 5th ESA Workshop on*, Noordwijk, Netherlands, Dec. 2010.
- [20] T. Paakki, J. Raasakka, F. Della Rosa, H. Hurskainen, and J. Nurmi, "TUTGNSS University based hardware/software GNSS receiver for research purposes," in *Ubiquitous Positioning Indoor Navigation and Location Based Service (UPINLBS), 2010*, Kirkkonummi, Finland, Oct. 2010.
- [21] S. Fantinato, L. Foglia, P. Iacone, D. Rovelli, C. Facchinetti, and A. Tuozzi, "PEGASUS-GNSS receiver platform for safety of life user segment," in *Satellite Navigation Technologies and European Workshop on GNSS Signals and Signal Processing,(NAVITEC), 2012 6th ESA Workshop on*, Noordwijk, Netherlands, Dec. 2012.
- [22] A. Fridman, and S. Semenov, "System-on-chip FPGA-based GNSS receiver," in *East-West Design & Test Symposium, 2013*, Rostov-on-Don, Russia, Sept. 2013.
- [23] G. Kappen, and T. G. Noll, "Application specific instruction processor based implementation of a GNSS receiver on an FPGA," in *Design, Automation and Test in Europe, 2006. DATE '06. Proceedings*, Munich, Germany, Mar. 2006.
- [24] M. S. Meraz, J. M. C. Arvizu, and A. J. A. Cruz, "GNSS receiver based on a SDR architecture using FPGA devices," in *Electronics, Robotics and Automotive Mechanics Conference (CERMA), 2011 IEEE*, Cuernavaca, Mexico, Nov. 2011.
- [25] M. Grondin, M. Belasic, L. Ries, J. L. Issler, P. Bataille, L. Jobey, and G. Richard, "A new operational low cost GNSS software receiver for microsatellites," in *Satellite Navigation Technologies and European Workshop on GNSS Signals and Signal Processing (NAVITEC), 2010 5th ESA Workshop on*, Noordwijk, Netherlands, Dec. 2010.
- [26] L. Ries, M. Monnerat, H. Al-Bitar, F. Legrand, M. Weyer, and G. Artaud, "Development of a flexible real time GNSS software receiver," in *Satellite Navigation Technologies and European Workshop on GNSS Signals and Signal Processing (NAVITEC), 2010 5th ESA Workshop on*, Noordwijk, Netherlands, Dec. 2010.
- [27] B. Sauriol, and R. Landry, "FPGA-based architecture for high throughput, flexible and compact real-time GNSS software defined receiver," in *Proceedings of the 2007 National Technical Meeting of The Institute of Navigation*, San Diego, CA, Jan. 2007, pp. 708-717.
- [28] J. Tian, W. Ye, S. Lin, and Z. Hua, "SDR GNSS receiver design over stand-alone generic TI DSP platform," in *Spread Spectrum Techniques and Applications, 2008 IEEE 10th International Symposium on*, Bologna, Italy, Aug. 2008.
- [29] T. E. Humphreys, M. L. Psiaki, P. M. Kintner Jr., and B. M. Ledvina, "GNSS Receiver Implementation on a DSP: Status, Challenges, and Prospects," in *Proceedings of the 19th International Technical Meeting of the Satellite Division of The Institute of Navigation (ION GNSS 2006)*, Fort Worth, TX, Sept. 2006, pp. 2370-2382.
- [30] J. Raasakka, H. Hurskainen, T. Paakki, and J. Nurmi, "Modeling multi-core software GNSS receiver with real time SW receiver," in *Proceedings of the 22nd International Technical Meeting of the Satellite Division of the Institute of Navigation (ION GNSS '09)*, vol. 1, Savannah, GA, USA, 2009, pp. 468–473.
- [31] E. G. Lightsey, T. E. Humphreys, J. A. Bhatti, A. J. Joplin, B. W. O'Hanlon, and S. P. Powell, "Demonstration of a space capable miniature dual frequency GNSS receiver," *Navigation, Journal of the Institute of Navigation*, vol. 61, no. 1, pp. 53–64, 2014.
- [32] S. Gleason, M. Quigley, and P. Abbeel, "Chapter 5: A GPS software receiver," in *GNSS Applications and Methods,* Norwood, MA, USA: Artech House, 2009, ch. 5, pp. 121-147.
- [33] S. A. Nik, and M. Petovello, "Multichannel dual frequency GLONASS software receiver," in *Proceedings of the 21st International Technical Meeting of the Satellite Division of the Institute of Navigation (ION GNSS '08)*, Sept. 2008, pp. 614–624.
- [34] M. G. Petovello, C. O'Driscoll, G. Lachapelle, D. Borio, and H. Murtaza, "Architecture and benefits of an advanced GNSS software receiver," *Journal of Global Positioning Systems*, vol. 7, no. 2, pp. 156-168, 2008.
- [35] S. Jeon, H. So, H. Noa, T. Lee, and C. Kee, "Development of real time software GPS receiver using Windows Visual C++ and USB RF front-end," in *Proceedings of the International Technical Meeting of the Institute of Navigation*, San Diego, CA, USA, 2010.
- [36] B. M. Ledvina, M. L. Psiaki, D. J. Sheinfeld, A. P. Cerruti, S. P. Powell, and P. M. Kintner Jr., "A real-time GPS civilian L1/L2 software receiver," in *Proceedings of the 17th International Technical Meeting of the Satellite Division of the Institute of Navigation (ION GNSS '04)*, Long Beach, CA, USA, Sept. 2004, pp. 986–1005.
- [37] F. Principe, C. Terzi, M. Luise, M. Casucci, "SOFT-REC: a GPS/EGNOS Software Receiver," in *14th IST Mobile & Wireless Communication Summit*, Dresden, Germany, 2005.
- [38] M. Anghileri, A. S. Ayaz, V. Kropp, J.-H. Won, B. Eissfeller, T. Pany, C. Stober, D. Dotterbock, I. Kramer, and D. S. Guixens, "ipexSR: A real-time multi-frequency software GNSS receiver," in *ELMAR, 2010 PROCEEDINGS*, Zadar, Croatia, Sept. 2010.
- [39] B. Huang, Z. Yao, F. Guo, S. Deng, X. Cui, and M. Lu, "STARx-a GPU based multi-system full-band real-time GNSS software receiver," in *Proceedings of the 26th International Technical Meeting of The Satellite Division of the Institute of Navigation (ION GNSS+ 2013)*, Nashville, TN, Sept. 2013, pp. 1549-1559.
- [40] A. Knezevic, C. O'Driscoll, and G. Lachapelle, "Co-processor aiding for real-time software GNSS receivers," in *Proceedings of the International Technical Meeting of the Satellite Division of the Institute of Navigation (ITM '10)*, vol. 2, San Diego, CA, USA, 2010, pp. 837–848.
- [41] G. Heinrichs, M. Restle, C. Dreischer, and T. Pany, "NavX-NSR - A Novel Galileo/GPS Navigation Software Receiver," in *Proceedings of the 20th International Technical Meeting of the Satellite Division of The Institute of Navigation (ION GNSS 2007)*, Fort Worth, TX, 2007, pp. 1329-1334.
- [42] T. E. Humphreys, M. Murrian, and L. Narula, "Low-cost precise vehicular positioning in urban environments," in *Proceedings of IEEE/ION PLANS 2018,* Monterey, CA, 2018, pp. 456-471.
- [43] J. Dampf, T. Pany, W. Bar, J. Winkel, C. Stober, K. Furlinger, P. Closas, and J. Garcia-Molina, "More than we ever dreamed possible: Processor technology for GNSS software receivers in the year 2015," *Inside GNSS*, vol. 10, no. 4, pp. 62–72, 2015.
- [44] P.-L. Normark and C. Stahlberg, "Hybrid GPS/Galileo real time software receiver," in *Proceedings of the 18th International Technical Meeting of the Satellite Division of The Institute of Navigation (ION GNSS 2005)*, Long Beach, CA, Sept. 2005, pp. 1906-1913.
- [45] L. L. Presti, P. di Torino, E. Falletti, M. Nicola, and M. T. Gamba, "Software defined radio technology for GNSS receivers," in *Metrology for Aerospace (MetroAeroSpace), 2014 IEEE*, Benevento, Italy, 2014, pp. 314-319.
- [46] A. Soghoyan, A. Suleiman, and D. Akopian, "A development and testing instrumentation for GPS software defined radio with fast FPGA prototyping support," *IEEE Trans. Instrum. Meas.*, vol. 63, no. 8, pp. 2001-2012, Feb. 2014.
- [47] K. Borre, D. Akos, N. Bertelsen, P. Rinder, and S. Jensen, *A Software-Defined GPS and Galileo Receiver: A Single-Frequency Approach.* Boston, MA: Birkhaauser, 2007.
- [48] F. Macchi, and M. G. Petovello, "Development of a one channel Galileo L1 software receiver and testing using real data," *in Proceedings of the 20th International Technical Meeting of the Satellite Division of the Institute of Navigation (ION GNSS '07)*, vol. 2, FortWorth, TX, USA, 2007, pp. 2256–2269.
- [49] D. F. M. Cristaldi, D. Margaria, and L. Lo Presti, "A multifrequency low-cost architecture for GNSS software receivers," in *Proceedings of the International Technical Meeting of the Satellite Division of the Institute of Navigation (ITM '10)*, vol. 2, San Diego, CA, USA, 2010, , pp. 849–857.
- [50] J. S. Silva, P. F. Silva, A. Fernandez, J. Diez, and J. F. M. Lorga, "Factored Correlator Model: A Solution for Fast, Flexible, and Realistic GNSS Receiver Simulations," in *Proceedings of the 20th International Technical Meeting of the Satellite Division of The Institute of Navigation (ION GNSS 2007)*, Fort Worth, TX, 2007, pp. 2676-2686.
- [51] Y Ng., G. X. Gao, "GNSS Multireceiver Vector Tracking," *IEEE Trans. Aerosp. Electron. Syst.*, vol. 53, no. 5, pp. 2583-2593, Oct. 2017.
- [52] C. Fernández-Prades, J. Arribas, and P. Closas, "GNSS-SDR: An Open Source Tool for Researchers and Developers," in *Proceedings of the 24th International Technical Meeting of The Satellite Division of the Institute of Navigation*, Portland, OR, Sept. 2011, pp. 780-794.
- [53] E. Schmidt, D. Akopian and D. J. Pack, "Development of a Real-Time Software-Defined GPS Receiver in a LabVIEW-Based Instrumentation Environment," *IEEE Trans. Instrum. Meas.*, vol. PP, no. 99, pp. 1-15, Mar. 2018, doi: 10.1109/TIM.2018.2811446.
- [54] E. Schmidt, and D. Akopian, "Exploiting Acceleration Features of LabVIEW Platform for Real-Time GNSS Software Receiver Optimization," in *Proceedings of the 30th International Technical Meeting of The Satellite Division of the Institute of Navigation (ION GNSS+ 2017)*, Portland, OR, Sep. 2017, pp. 3694-3709.

[55] F. Principe, G. Bacci, F. Giannetti, and M. Louise, "Software-defined radio technologies for GNSS receivers: A tutorial approach to a simple design and implementation," *Int. J. Navig. Obs.*, 2011, doi:10.1155/2011/979815.

[56] G. Hein, J.-H. Won, and T. Pany, "GNSS software defined radio: real receiver or just a tool for experts?" *Inside GNSS*, vol. 1, no. 5, 2006.

[57] D. Akopian, "Fast FFT based GPS satellite acquisition methods," *IEE Proceedings-Radar, Sonar and Navigation*, vol. 152, no. 4, pp. 277-286, Aug. 2005.

[58] E. D. Kaplan, *Understanding GPS, Principles and Applications*, Boston, MA: Artech House, 1996.

[59] T. E. Humphreys, J. Bhatti, T. Pany, B. Ledvina, and B. O'Hanlon, "Exploiting multicore technology in software-defined GNSS receivers," in *Proceedings of the ION GNSS Meeting, (Savannah, GA)*, Savannah, GA, 2009, pp. 326–338.

[60] The GNU Radio Foundation, Inc, "GNU Radio," [Online]. Available: www.gnuradio.org. [Accessed: January 2015].

[61] E. Falletti, B. Motella, and M. T. Gamba, "Post-correlation signal analysis to detect spoofing attacks in GNSS receivers," in *Signal Processing Conference (EUSIPCO), 2016 24th European*, Budapest, Hungary, 2016.

[62] A. Broumandan, A. Jafarnia-Jahromi, and G. Lachapelle, "Spoofing detection, classification and cancelation (SDCC) receiver architecture for a moving GNSS receiver," *GPS Solutions*, vol. 19, pp. 475-487, 2015.

[63] A. Ranganathan, H. Olafsdottir, and S. Capkun, "SPREE: A Spoofing Resistant GPS Receiver," in *Proceedings of the 22nd Annual International Conference on Mobile Computing and Networking*, New York, NY, USA, Oct. 3-7, 2016, pp. 348-360.

[64] M. Paonni, J. G. Jang, B. Eissfeller, S. Wallner, J. A. A. Rodriguez, J. Samson, and F. A. Fernandez, "Innovative interference mitigation approaches: analytical analysis, implementation and validation," in *Satellite Navigation Technologies and European Workshop on GNSS Signals and Signal Processing (NAVITEC), 2010 5th ESA Workshop on*, Noordwijk, Netherlands, Feb. 2010.

[65] A. Jafarnia-Jahromi, T. Lin, A. Broumandan, J. Nielsen, and G. Lachapelle, "Detection and Mitigation of Spoofing Attacks on a Vector Based Tracking GPS Receiver," in *Proceedings of International Technical Meeting of the Institute of Navigation (ION ITM 2012)*, Newport Beach, CA, Jan. 2012, pp. 790-800.

[66] Y. Ng, and G. X. Gao, "Mitigating jamming and meaconing attacks using direct GPS positioning," in *Position, Location and Navigation Symposium (PLANS), 2016 IEEE/ION*, Savannah, GA, USA, May 2016.

[67] Y. Ng, and G. X. Gao, "Robust GPS-based direct time estimation for PMUs," in *Position, Location and Navigation Symposium (PLANS), 2016 IEEE/ION*, Savannah, GA, USA, May 2016.

[68] E. Schmidt, Z. A. Ruble, D. Akopian, and D. J. Pack "A Reduced Complexity Cross-correlation Interference Mitigation Technique on a Real-Time Software-defined Radio GPS L1 Receiver," in *Proceedings of IEEE/ION PLANS 2018*, Monterey, CA, 2018, pp. 931-939.

[69] W. L. Myrick, M. Picciolo, J. S. Goldstein, and V. Joyner, "Multistage anti-spoof GPS interference correlator (MAGIC)," in *Military Communications Conference, MILCOM 2015-2015 IEEE*, Tampa, FL, USA, Oct. 2015.

[70] T. E. Humphreys, "Detection Strategy for Cryptographic GNSS Anti-Spoofing," *IEEE Trans. Aerosp. Electron. Syst.*, vol. 49, no. 2, pp. 1073-1090, Apr. 2013.

[71] T. E. Humphreys, B. M. Ledvina, M. L. Psiaki, B. W. O'Hanlon, and P. M. Kintner Jr., "Assessing the spoofing threat: Development of a portable GPS civilian spoofer," in *Proceedings of the 21st International Technical Meeting of the Satellite Division of The Institute of Navigation (ION GNSS 2008)*, Savannah, GA, Sept. 2008, pp. 2314-2325.

[72] T. E. Humphreys, D. P. Shepard, J. A. Bhatti, and K. D. Wesson, "A testbed for developing and evaluating GNSS signal authentication techniques," in *Proc. Int. Symp. Certif. GNSS Syst. Serv. (CERGAL)*, Dresden, Germany, July 2014.

[73] W. O'Hanlon, M. L. Psiaki, J. A. Bhatti, D. P. Shepard, and T. E. Humphreys, "Real-Time GPS Spoofing Detection via Correlation of Encrypted Signals," in *Navigation*, vol. 60, pp. 267-278, 2013.

[74] M. L. Psiaki, B. W. OHanlon, J. A. Bhatti, D. P. Shepard, and T. E. Humphreys, "GPS spoofing detection via dual-receiver correlation of military signals," *IEEE Trans. Aerosp. Electron. Syst.*, vol. 49, pp. 2250-2267, 2013.

[75] W. O'Hanlon, M. L. Psiaki, T. E. Humphreys, and J. A. Bhatti, "Real-time spoofing detection in a narrow-band civil GPS receiver," in *Proceedings of the 23rd International Technical Meeting of The Satellite Division of the Institute of Navigation (ION GNSS 2010)*, Portland, OR, Sept. 2010, pp. 2211-2220.

[76] W. O'Hanlon, M. L. Psiaki, T. E. Humphreys, and J. A. Bhatti, "Real-time spoofing detection using correlation between two civil GPS receiver," in *Proceedings of the 25th International Technical Meeting of The Satellite Division of the Institute of Navigation (ION GNSS 2012)*, Nashville, TN, Sept. 2012, pp. 3584-3590.

[77] L. Heng, J. J. Makela, A. D. Dominguez-Garcia, R. B. Bobba, W. H. Sanders and G. X. Gao, "Reliable GPS-based timing for power systems: A multi-layered multi-receiver architecture," in *Power and Energy Conference at Illinois (PECI), 2014*, 2014.

[78] T. E. Humphreys, J. A. Bhatti, D. P. Shepard, and K. D. Wesson, "The Texas spoofing test battery: Toward a standard for evaluating GPS signal authentication techniques," in *Proceedings of the 25th International Technical Meeting of The Satellite Division of the Institute of Navigation (ION GNSS 2012)*, Nashville, TN, September 2012, pp. 3569-3583.

[79] S. Daneshmand, A. Jafarnai-Jahromi, A. Broumandan, J. Nielsen, and G. Lachapelle, "GNSS spoofing mitigation in multipath environments using space-time processing," in *European navigation conference (ENC) 2013*, 2013.

[80] M. L. Psiaki, B. W. O'Hanlon, S. P. Powell, J. A. Bhatti, K. D. Wesson, and T. E. Humphreys, "GNSS spoofing detection using two-antenna differential carrier phase," in *Proceedings of the 27th International Technical Meeting of The Satellite Division of the Institute of Navigation (ION GNSS+ 2014)*, Tampa, FL, Sept. 2014, pp. 2776-2800.

[81] P. Y. Montgomery, T. E. Humphreys, and B. M. Ledvina, "Receiver-autonomous spoofing detection: Experimental results of a multi-antenna receiver defense against a portable civil GPS spoofer," in *Proceedings of the 2009 International Technical Meeting of The Institute of Navigation*, Anaheim, CA, Jan. 2009, pp. 124-130.

[82] S. Daneshmand, A. Jafarnia-Jahromi, A. Broumandan, and G. Lachapelle, "A GNSS structural interference mitigation technique using antenna array processing," in *Sensor Array and Multichannel Signal Processing Workshop (SAM), 2014 IEEE 8th*, Coruna, Spain, Aug. 2014.

[83] S. Daneshmand, A. Jafarnia-Jahromi, A. Broumandan, and G. Lachapelle, "A low-complexity GPS anti-spoofing method using a multi-antenna array," in *Proceedings of the 25th International Technical Meeting of The Satellite Division of the Institute of Navigation (ION GNSS 2012)*, Nashville, TN, Sept. 2012, pp. 1233-1243.

[84] A. Jafarnia-Jahromi, A. Broumandan, S. Daneshmand, N. Sokhandan, and G. Lachapelle, "A double antenna approach toward detection, classification and mitigation of GNSS structural interference," in *Proceedings of NAVITEC 2014*, Noordwijk, Netherlands, Dec. 2014.

[85] C. Fernandez-Prades, J. Arribas, P. Closas, "Robust GNSS Receivers by Array Signal Processing: Theory and Implementation." *Proc. IEEE*, vol. 104, no. 6, pp. 1207-1220, Jun. 2016.

[86] J. Arribas, C. Fernandez-Prades, P. Closas, "Multi-antenna techniques for interference mitigation in GNSS signal acquisition," *P. EURASIP J. Adv. Signal Process*, vol. 2013, no. 1, Sep. 2013.

[87] J. Arribas, P. Closas, and C. Fernández-Prades, "Interference Mitigation in GNSS Receivers by Array Signal Processing: A Software Radio Approach," in *Proceedings of the 8th IEEE Sensor Array and Multichannel Signal Processing Workshop*, A Coruna, Spain, Jun. 2014, pp. 121-124.

[88] Ettus Research, a National Instruments (NI) company, "Ettus Research - The leader in Software Defined Radio (SDR)," [Online]. Available: https://www.ettus.com/. [Accessed: January 2015].

[89] NSL Primo GNSS SDR Front End. [Online]. Available: http://www.nsl.eu.com/primo.html. [Accessed: January 2015].

[90] SX3 GNSS Software Receiver. [Online]. Available: http://www.ifen.com. [Accessed: January 2015].

[91] Intel Corporation, "Intel(R) C++ Intrinsic Reference," [Online]. Available: https://software.intel.com/sites/default/files/a6/22/18072-347603.pdf. [Accessed: June 2017].

[92] M. Frigo, and S. G. Johnson, "The Design and Implementation of FFTW3," *Proc. IEEE*, vol. 93, no. 2, pp. 216-231, 2005.

[93] Eigen, "Eigen - a C++ template library for linear algebra: matrices, vectors, numerical solvers, and related," [Online]. Available: http://eigen.tuxfamily.org. [Accessed: 2016].

[94] U. Madhow, and M. L. Honig, "MMSE Interference Suppression for Direct-Sequence Spread-Spectrum CDMA," *IEEE Trans. Commun.*, vol. 42, no. 12, pp. 3178-3188, 1994.

[95] R. Kohno, H. Imai, M. Hatori, and S. Pasupathy, "An Adaptive Canceller of Cochannel Interference for Spread-Spectrum Multiple-Access

Communication Networks in a Power Line," *IEEE J. Sel. Areas Commun.*, vol. 8, no. 4, pp. 691-699, 1990.
[96] G. Lopez-Risueno, and G. Seco-Granados, "CN0 Estimation and Near-Far Mitigation for GNSS Indoor Receivers," in *Vehicular Technology Conference, 2005. VTC 2005-Spring. 2005 IEEE 61st*, Stockholm, Sweden, Dec. 2005.
[97] A. G. Dempster, and E. P. Glennon, "Apparatus and Method for Mitigation of Cross Correlation in GPS System," U.S. Patent 20070058696, Mar. 15., 2007.
[98] U. Madhow, M. L. Honig, and S. Verdu, "Blind Adaptive Multiuser Detection," *IEEE Trans. Inf. Theory*, vol. 41, no. 4, pp. 944-960, 1995.
[99] L. Rugini, P. Banelli, and S. Cacopardi, "A Full-Rank Regularization Technique for MMSE Detection in Multiuser CDMA Systems," *IEEE Commun. Lett.*, vol. 9, no. 1, pp. 34-36, 2005.
[100] G. H. Golub, and C. F. Van Loan, *Matrix Computations*, 3rd ed., Baltimore, Maryland: The Johnson Hopkins University Press, 1996.
[101] National Instruments, "NI Global Navigation Satellite System Toolkits – National Instruments" [Online]. Available: http://sine.ni.com/nips/cds/view/p/lang/en/nid/204980. [Accessed: 2016].
[102] M. A. Abu-Rgheff, "Pseudo-Random Code Sequences for Spread-Spectrum Systems," in *Introduction to CDMA Wireless Communications*, London, UK: Academic Press, 2007, pp. 203-220.

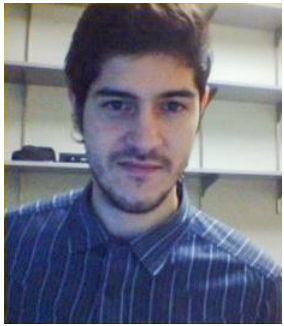

**Erick Schmidt** (S'17) received the B.Sc. degree in electronics and computer engineering with Honors from The Monterrey Institute of Technology and Higher Education, Monterrey, Mexico, in 2011, and the M.Sc. degree from the University of Texas at San Antonio, San Antonio, TX, USA, in 2015. From 2011 to 2013, he was a Systems Engineer with Qualcomm Incorporated.

He is currently working towards the Ph.D. degree in electrical engineering from the University of Texas at San Antonio, San Antonio, TX, USA. His research interests include software-defined radio, indoor navigation, global navigation satellite system (GNSS), and fast prototyping algorithms and accelerators for baseband communication systems. He is a student member of the IEEE and the Institute of Navigation (ION).

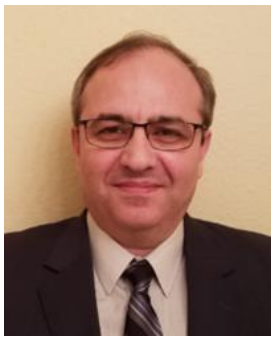

**David Akopian** (M'02-SM'04) is a Professor at the University of Texas at San Antonio (UTSA). Prior to joining UTSA, he was a Senior Research Engineer and Specialist with Nokia Corporation from 1999 to 2003. From 1993 to 1999 he was a researcher and instructor at the Tampere University of Technology, Finland, where he received his Ph.D. degree in 1997. Dr. Akopian's current research interests include digital signal processing algorithms for communication and navigation receivers, positioning, dedicated hardware architectures and platforms for software defined radio and communication technologies for healthcare applications.

He authored and co-authored more than 30 patents and 140 publications. He is elected as a Fellow of US National Academy of Inventors in 2016. He served in organizing and program committees of many IEEE conferences and co-chairs an annual conference on Multimedia and Mobile Devices. His research has been supported by National Science Foundation, National Institutes of Health, USAF, US Navy, and Texas foundations.

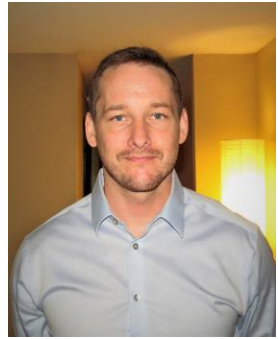

**Zachary Ruble** received the B.S. degree in computer engineering from the University of Wyoming, Laramie, WY USA in 2007 and the M.S. and Ph.D. degrees in electrical engineering also at the University of Wyoming in 2009 and 2015 respectively. He is currently a postdoctoral fellow in the Unmanned Systems Laboratory at the University of Tennessee at Chattanooga, Chattanooga, TN, USA.

From 2010 to 2015, he was a Research Assistant with the Electrical and Computer Engineering Dept. at the University of Wyoming. Since 2015, he has been with the Unmanned Systems Laboratory at the University of Tennessee Chattanooga. His research interest includes cooperative unmanned systems, robotics, satellite systems, navigation, and distributed control systems.

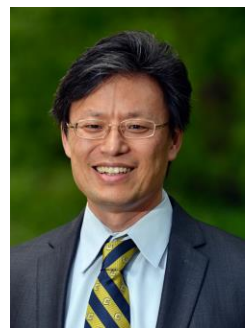

**Daniel J. Pack** received the B.S. degree in computer engineering from the University of Wyoming, Laramie, WY USA in 2007 and the M.S. and Ph.D. degrees in electrical engineering also at the University of Wyoming in 2009 and 2015 respectively. He is currently a postdoctoral Fellow in the Unmanned Systems Laboratory at the University of Tennessee at Chattanooga, Chattanooga, TN, USA.

From 2010 to 2015, he was a Research Assistant with the Electrical and Computer Engineering Dept. at the University of Wyoming. Since 2015, he has been with the Unmanned Systems Laboratory at the University of Tennessee Chattanooga. His research interest includes cooperative unmanned systems, robotics, embedded systems, and distributed control systems.

degree in electrical engineering from Purdue University, West Lafayette, IN, USA, in 1995.

He is currently the Dean of the College of Engineering and Computer Science, University of Tennessee, Chattanooga

(UTC), TN, USA. Prior to joining UTC, he was a Professor and the Mary Lou Clarke Endowed Chair of the Electrical and Computer Engineering Department, University of Texas, San Antonio and a Professor (now Professor Emeritus) of electrical and computer engineering at the United States Air Force Academy, CO, USA, where he was the founding Director of the Academy Center for Unmanned Aircraft Systems Research. His research interests include unmanned aerial vehicles, intelligent control, automatic target recognition, robotics, and engineering education.